\documentclass[12pt]{article}
\usepackage{amsmath,amssymb,amsfonts,amsthm}
\usepackage[english]{babel}
\usepackage{graphicx}
\usepackage{float}
\newcommand{\be}{\begin{equation}}
\newcommand{\ee}{\end{equation}}
\newcommand{\bea}{\begin{eqnarray}}
\newcommand{\eea}{\end{eqnarray}}
\newcommand{\nn}{\nonumber \\}
\newcommand{\p}[1]{(\ref{#1})}
\newcommand{\lb}{\label}

\usepackage{caption}
\usepackage{subcaption}

\begin{document}
\begin{titlepage}
\rightline{JINR E2-2013-78}

\vfill

\begin{center}
\baselineskip=16pt {\Large\bf
Deformed Supersymmetric Mechanics}

\vskip 0.3cm {\large {\sl }} \vskip 10.mm {\bf $\;$  E. Ivanov$^{\,a}$,  $\;$ S. Sidorov$^{\,b}$
}
\vspace{1cm}

{\it Bogoliubov Laboratory of Theoretical Physics, JINR, \\
141980 Dubna, Moscow Region, Russia\\
}
\end{center}
\vfill

\par
\begin{center}
{\bf Abstract}
\end{center}
Motivated by a recent interest in curved rigid supersymmetries, we construct a new type of ${\cal N}{=}4$, $d{=}1$ supersymmetric systems by employing superfields
defined on the cosets of the supergroup $SU(2|1)$.
The relevant worldline supersymmetry is a deformation of the standard ${\cal N}{=}4, d{=}1$ supersymmetry by a mass parameter $m$.
As instructive examples we consider, at the classical and quantum levels,  the models associated with the supermultiplets ({\bf 1,4,3}) and ({\bf 2,4,2})
and find out interesting interrelations
with some previous works on non-standard $d{=}1$  supersymmetry. In particular,  the $d{=}1$ systems with ``weak supersymmetry'' are naturally
reproduced within our  $SU(2|1)$  superfield approach as a subclass of the ({\bf 1,4,3}) models. A generalization to the ${\cal N}{=}8, d{=}1$ case
implies the supergroup $SU(2|2)$ as the candidate deformed worldline supersymmetry.
\vspace{2cm}

\noindent PACS: 03.65-w, 11.30.Pb, 04.60.Ds, 02.40.Tt \\
\noindent Keywords: supersymmetry, superfields, deformation

\begin{quote}
\vfill \vfill \vfill \vfill \vfill \hrule width 5.cm \vskip 2.mm
{\small
\noindent $^a$ eivanov@theor.jinr.ru\\
\noindent $^b$ sidorovstepan88@gmail.com\\
}
\end{quote}
\end{titlepage}

\setcounter{footnote}{0}

\setcounter{page}{1}

\numberwithin{equation}{section}

\section{Introduction}
Recently, there was a growth of  interest in rigid supersymmetric theories in diverse dimensions, such that the relevant supersymmetry groups include,
as the bosonic subgroups, the groups of motion of some curved spaces (see, e.g., \cite{FZ,DFZ}). This should be contrasted with the standard rigid supersymmetric
theories in which the bosonic invariance subgroup is the Poincar\'e group, the group of motion of the flat Minkowski space. There is the hope that
the study of the new class of theories will give rise to a further progress in understanding the generic gauge/gravity correspondence.

The simplest Poincar\'e supergroup is the $d=1$ one,
\be
\{Q^m, Q^n\} = 2 \delta^{mn}\,H\,, \quad [H, Q^m] = 0\,, \; m = 1, \ldots {\cal N}\,,\lb{Nsusy}
\ee
where $Q^m$ are ${\cal N}$ real supercharges and $H$ is the time-translation generator. The associated systems are models
of supersymmetric quantum mechanics (SQM) \cite{W1}, with $H$ being the relevant Hamiltonian. The SQM models, including their versions with
extended ${\cal N}>2$ supersymmetry, have a lot of applications in various physical and mathematical domains. It is tempting to construct SQM models
based on some curved versions of the $d=1$ Poincar\'e supersymmetry. They could be considered as the $d=1$ analogs of the higher-dimensional
supersymmetric models just mentioned and, in some cases, follow from the latter via dimensional reduction.
Irrespective of the dimensional reduction reasoning, they can bear an obvious interest on their own as non-trivial self-consistent deformations
of the standard SQM models with plenty of potential applications.

One possible way to define such generalized SQM models is suggested by the simplest non-trivial $d=1$ Poincar\'e
superalgebra, the ${\cal N}=2$ one. Introducing complex generators
$$
Q = \frac{1}{\sqrt{2}}(Q^1 + i Q^2)\,, \quad \bar Q = \frac{1}{\sqrt{2}}(Q^1 - i Q^2)\,,
$$
the superalgebra \p{Nsusy} for ${\cal N}=2$ can be rewritten as
\be
\{Q, \bar Q\} = 2 H\,, \quad Q^2 = \bar Q^2 = 0, \quad [H, Q] = [H, \bar Q] = 0\,.\lb{N2}
\ee
It is instructive to add the commutators with the generator $J$ of the group $O(2) \sim U(1)$ which is the automorphism group of \p{Nsusy} for
${\cal N}=2$:
\be
[J, Q] = Q\,, \quad [J, \bar Q] = - \bar Q\,, \quad [H, J] = 0\,.\lb{JQ}
\ee

On the one hand, the relations \p{N2} and \p{JQ} define the ${\cal N}=2, d=1$ Poincar\'e superalgebra. On the other hand (and this fact is less known),
these relations are recognized as defining the superalgebra $u(1|1)$, with $H$ being the relevant central charge generator. After factoring out the $U(1)$
generator $J$, we are left with the superalgebra $su(1|1)$ \p{N2}.

This twofold interpretation of ${\cal N}=2, d=1$ Poincar\'e superalgebra suggests two ways of extending it to higher-rank $d=1$ supersymmetries.

The first one is the straightforward extension
\be
({\cal N}=2\,, \; d=1) \quad \Rightarrow \quad ({\cal N} > 2\,, \; d=1\;\mbox{Poincar\'e})\,,\lb{stand}
\ee
where the general ${\cal N}, d=1$ Poincar\'e superalgebra is defined by the relations \p{Nsusy}. Except for ${\cal N}=2$, these algebras
cannot be identified
with any simple or semi-simple superalgebras (though can still be recovered through contractions and/or truncations of such superalgebras).
The possible extra bosonic
generators are those of the automorphism group (it is $O({\cal N})$ in the general case) and/or central (or ``semi-central'')
charge generators which commute with
the supercharges.

Another, less evident opportunity corresponds to the following chain of embeddings
\be
({\cal N}=2\,, \; d=1) \equiv u(1|1) \; \subset \; su(2|1) \; \subset \; su(2|2) \; \subset \; \ldots\,.\lb{nonstand}
\ee
The characteristic feature of this sort of extensions is that the relevant superalgebras necessarily contain, besides an analog
of the Hamiltonian $H$, also
additional bosonic generators which form some internal symmetry subgroups commuting with the ``would-be'' Hamiltonian. They appear in the closure
of the supercharges,  and {\it do not commute} with the latter (as opposed, e.g.,  to the central charges in the Poincar\'e superalgebras).
Though the chain \p{nonstand} is certainly non-unique, in the sense that one could imagine some other extensions of $u(1|1)$ among its links,
the superalgebras written down  in \p{nonstand} are distinguished in that they seem to be minimal deformations of the ${\cal N}=4$ and ${\cal N}=8$
one-dimensional Poincar\'e superalgebras: they go over into the latter, when taking the contraction limit with respect to some dimensionful parameter.

The supergroup $SU(2|1)$ as the alternative of the standard worldline  ${\cal N}=4, d=1$ supersymmetry in SQM models already appeared in literature under
the name ``weak supersymmetry'' \cite{WS} (see also \cite{BN3,BN4}), though no explicit identification of the latter with $SU(2|1)$
was made\footnote{A complexified version of $SU(2|1)$ as a hidden symmetry of some  ${\cal N}=2$ SQM model with higher derivatives (and ghosts)
was found in \cite{SRo}.} and
no systematic methods of constructing such new SQM models were given. One of the basic aims of the present paper is to develop such methods,
which would be applicable not only to $SU(2|1)$, but also to the case of the supergroup $SU(2|2)$ and, hopefully, to other interesting
examples of this type.

We construct the worldline superfield approach to $SU(2|1)$ and demonstrate that most of the
off-shell multiplets of ${\cal N}=4, d=1$ supersymmetry have the well-defined $SU(2|1)$ analogs. In particular, the models considered in \cite{WS}
are based on the $SU(2|1)$ multiplet ${\bf (1, 4, 3)}\,$, and we reproduce these models from our superfield approach.  Some peculiarities of their
quantum spectra find a natural explanation in the framework of the $SU(2|1)$ representation theory \cite{repres}, based
on the property that the relevant Casimir operators have a notable expression in terms of the Hamiltonian. This supergroup
has also invariant chiral subspaces which are natural carriers of the chiral
superfields encompassing off-shell multiplets ${\bf (2, 4, 2)}$, for which we also construct general superfield and component actions.
An interesting new feature of these actions is the inevitable presence of the bosonic $d=1$ Wess-Zumino terms of the first order in time derivative,
parallel with the standard second-order kinetic terms. We also show that $SU(2|1)$ admits a supercoset which is an analog of the harmonic
analytic superspace of
the standard ${\cal N}=4, d=1$ supersymmetry \cite{ILech}. This means that one can define $SU(2|1)$ analogs of the ``root'' ${\cal N}=4$ multiplet
$({\bf 4, 4, 0})$ and of the multiplet $({\bf 3, 4, 1})$ by embedding them into the appropriate harmonic analytic superfields.
Detailed analysis of these and related issues (including, e.g.,  the appropriate generalization of the $d=1$ superfield gauging procedure \cite{DI})
will be given elsewhere.

The paper is organized as follows. The $SU(2|1)$ superspace is
constructed in Section 2. The study of the $SU(2|1)$ SQM models
based on the multiplet $({\bf 1, 4, 3})$ is performed in Sections 3
and 4. The similar study for the $({\bf 2, 4, 2})$ multiplets is the
subject of Sections 5 and 6. The summary and some problems for the
future analysis are the contents of Section 7. In Appendix A, some
special cases of the quantum ${\bf (2, 4, 2)}$ models are
considered. In Appendix B we briefly treat
the $({\bf 1, 4, 3})$ and $({\bf 2, 4, 2})$ models in the equivalent ${\cal N}=2$, $d=1$ superfield language
and show that they supply examples of deformed ${\cal N}=2$ supersymmetric mechanics.

\setcounter{equation}{0}
\section{$SU(2|1)$ superspace}

\subsection{The $su(2|1)$ algebra}
We start with the following form of the (centrally-extended) superalgebra $su(2|1)$:
\bea
    &&\lbrace Q^{i}, \bar{Q}_{j}\rbrace = 2m\left( I^i_j -\delta^i_j F\right)+ 2\delta^i_j H\,,\qquad\left[I^i_j,  I^k_l\right]
    = \delta^k_j I^i_l - \delta^i_l I^k_j\,,\nn
    &&\left[I^i_j, \bar{Q}_{l}\right] = \frac{1}{2}\delta^i_j\bar{Q}_{l}-\delta^i_l\bar{Q}_{j}\, ,\qquad \left[I^i_j, Q^{k}\right]
    = \delta^k_j Q^{i} - \frac{1}{2}\delta^i_j Q^{k},\nn
    &&\left[F, \bar{Q}_{l}\right]=-\frac{1}{2}\bar{Q}_{l}\,,\qquad \left[F, Q^{k}\right]=\frac{1}{2}Q^{k}.\label{algebra}
\eea
All other (anti)commutators are vanishing.

The dimensionless generators $I^i_j$ and $F$ generate $U(2)$ symmetry, while the mass-dimension generator $H$ commutes
with everything and so can be interpreted as the central charge generator. In the quantum-mechanical realization of $SU(2|1)$ we will be  interested in,
this generator becomes just the canonical Hamiltonian, while in the superspace realization it is interpreted as the time-translation generator.
The mass parameter $m$ is arbitrary and
it is introduced in order to separate the generator $H$ from the internal symmetry generator $F$ which possesses non-trivial commutation relations
with the fermionic
generators. It can be treated as the contraction parameter: sending $m\,\rightarrow \,0$  leads to the standard ${\cal N}=4, d=1$ Poincar\'e superalgebra.
In the limit $m=0\,$, the generators $I^i_j$ and $F$ become the $U(2)$ automorphism generators of this ${\cal N}=4, d=1$ superalgebra.
It is worth mentioning that
the full automorphism group of the flat ${\cal N}=4, d=1$ is $SO(4) \sim SU(2)\times SU(2)$ and, after reduction, $F$ is recognized as belonging
just to the second $SU(2)$ factor in this product. At $m\neq 0\,$, only the generator $F$ is present,
as the only counterpart of the second automorphism $SU(2)$ of the $m=0$ case.

The central extension \p{algebra} is in fact isomorphic to the semi-direct sum of the genuine $su(2|1)$ superalgebra (without central charge) and
an external automorphism ($R$-symmetry) $U(1)$ generator \cite{FZ}. Passing to the new basis
in \p{algebra} as $(H, F) \rightarrow ({F}' = F - \frac{1}{m}H, F)\,$, we observe that in this basis
the generator ${F}'$ becomes the internal $U(1)$ generator which, together with $Q^i, \bar Q_i, I^i_j$,
form the centerless $su(2|1)$ algebra, while $F$ becomes the outer automorphism $U(1)$ generator possessing
the same commutation relations with the supercharges as in \p{algebra}. Despite this difference,
in what follows we will refer to \p{algebra} as the $su(2|1)$ superalgebra, hoping
that this will not give rise to any confusion.

Note that the substitutions
\be
m\rightarrow -m\,,\; Q^i \rightarrow \bar{Q}^i\,,\;  \bar{Q}_j\rightarrow -Q_j\,,\;
I^i_j\rightarrow \varepsilon_{jk}\varepsilon^{il}I^k_l\,,\;H\rightarrow H\,,\;
F\rightarrow -F  \label{reflection}
\ee
leave the superalgebra \p{algebra} intact, i.e. they constitute an automorphism of \p{algebra}. This means that the cases
of positive and negative $m$ are in fact
equivalent, and so in what follows we can limit our consideration to $m>0\,$.

\subsection{Coset superspace}
The supergroup $SU(2|1)$ can be realized by left shifts on a few coset supermanifolds. The supercosets which have appeared so far  in diverse variants
of the super Landau problem with $SU(2|1)$ as the target space supersymmetry include $SU(2|1)/U(1|1)$ ($({\bf 2|2})$ dimensional supersphere,
with the sphere
$S^2$ as the bosonic submanifold)\cite{ssph},
$SU(2|1)/[U(1)\times U(1)]$ ($({\bf 2|4})$ dimensional superflag, again with $S^2$ as the bosonic submanifold)\cite{sflag} and $SU(2|1)/U(2)$
(purely odd coset of the dimension $({\bf 0|4})$)\cite{04cos,04cos1}. One could also consider, e.g., the supercoset  $SU(2|1)/U(1)$ with $S^3$
as the bosonic submanifold
and the full $SU(2|1)$ group manifold as a superextension of $S^1\times S^3$ (or $R^1\times S^3$).
In all these realizations the coset parameters are regarded as some worldline fields, in accordance with the treatment of $SU(2|1)$
as a nonlinearly realized internal supersymmetry. The relevant Hamiltonians are purely external: they commute with all $SU(2|1)$ generators, but
never come out in the closure of the latter.

Here we will be interested in the $SU(2|1)$ coset of the entirely different type. It is a direct  analog of the standard ${\cal N}=4, d=1$
superspace \cite{ILech,ikl1}, with the coset parameters being identified with the coordinates, not with the fields. The fields will finally appear
as the components of the appropriate superfields given on this supercoset. The splitting of the $U(2)$ singlet generator in \p{algebra} into
the $H$ and $F$  parts plays the crucial role for the possibility to define such a coset supermanifold in the self-consistent way. We place
the $U(2)$ generators into the stability subgroup and are left with $H$, $Q_i$ and $\bar Q^i$ as the coset generators
\be
    \frac{SU(2|1)}{SU(2)\times U(1)}\, \sim \,\frac{\{Q^{i},\bar{Q}_{j},H,I^i_j,F \}}{\{I^i_j,F \}}\,.
\ee
The corresponding superspace coordinates $\{t,\theta_{i}, \bar{\theta}^{j}\}$ are then identified with the parameters
associated with these coset generators. An element of this supercoset can be conveniently parametrized as
\be
g= \exp{\left( i t H+ i\tilde\theta_{i}Q^{i}-i\tilde{\bar{\theta}}^{j}\bar{Q}_{j}\right)}\,,
\ee
where\footnote{We use the convention $\left(\bar{\theta}\cdot\theta\right)= \bar{\theta}^k\theta_k$}
\bea
\tilde\theta_i =\left[1-\frac{2m}{3}\left(\bar{\theta}\cdot\theta\right)\right]\theta_i\,.
\eea
\subsection{Cartan forms}
Prior to giving how $SU(2|1)$ is realized on the superspace coordinates, it is convenient to calculate the left-covariant Cartan one-forms. They
are defined by the standard relation
\be
\Omega :=    g^{-1}d g =e^{-B}d\, e^{B} +i\, dt \,H=i\Delta\theta_{i}Q^{i}-i\Delta\bar{\theta}^{j}\bar{Q}_{j}
+i\Delta h^j_i\,I^i_j+i\Delta \hat{h}\,F +i \Delta t \,H\,,\lb{C1}
\ee
where
\be
    B := i\left(\tilde\theta_{i}Q^{i}-\tilde{\bar{\theta}}^{j}\bar{Q}_{j}\right). \lb{Bdef}
\ee
Using the nilpotency property of the fermionic coordinates and the (anti)commutation relations \p{algebra},
it is straightforward to find the explicit expressions for the Cartan forms
\bea
    \Delta\theta_{i}&=&d\theta_{i}+{m}\left( d\theta_{l} \bar{\theta}^{l}\theta_{i}
    - d\theta_{i} \bar{\theta}^{k}\theta_{k}\right)+\frac{m^2}{4}d\theta_{i} \left(\bar{\theta}\cdot\theta\right)^2,\nn
    \Delta\bar{\theta}^{j}&=&d\bar{\theta}^{j}- {m}\left( d\bar{\theta}^{l}\theta_{l}\bar{\theta}^{j}
    -d\bar{\theta}^{j} \theta_{k}\bar{\theta}^{k}\right)+\frac{m^2}{4}d\bar{\theta}^{j} \left(\bar{\theta}\cdot\theta\right)^2,\nn
    \Delta t&=& d t + i\left( d\theta_{i} \bar{\theta}^{i}+d\bar{\theta}^{i} \theta_{i}\right)
    \left[1- 2m\left(\bar{\theta}\cdot\theta\right)\right],\nn
    \Delta \hat{h}&=& -{im}\left( d\theta_{i} \bar{\theta}^{i}+d\bar{\theta}^{i} \theta_{i}\right)
    \left[1-2m\left(\bar{\theta}\cdot\theta\right)\right],\nn
    \Delta h^j_i &=& {im}\left[ d\theta_{i} \bar{\theta}^{j}+d\bar{\theta}^{j} \theta_{i}
    -\frac12{\delta^j_i} \left(d\theta_{l} \bar{\theta}^{l} + d\bar{\theta}^{l}\theta_{l}\right)\right]
    \left[1-\frac{3m}{2}\left(\bar{\theta}\cdot\theta\right)\right]\nn
    && -\frac{i m^2}{2} \left(d\theta_{l} \bar{\theta}^{l} + d\bar{\theta}^{l}\theta_{l}\right)
    \left[ \bar{\theta}^{j}\theta_{i} - \frac12{\delta^j_i} \left(\bar{\theta}\cdot\theta\right)\right]. \lb{C2}
\eea

\subsection{Transformation properties}
The transformation properties of the superspace coordinates under the left shifts with the parameters $\epsilon_{i}$ and $\bar{\epsilon}^{j}$,
as well as the induced infinitesimal transformations belonging to the stability subgroup $(I^i_j,F)$, can be found from the general formula
\be
\left(1+ i\epsilon_{i}Q^{i}-i\bar{\epsilon}^{i}\bar{Q}_{i}\right)g = g{\,}'\, h\,,\lb{Tr1}
\ee
where
\be
h = 1+\left(i\delta h^j_i\,I^i_j+i\delta \hat{h}\, F\right). \lb{Tr2}
\ee
Eqs. \p{Tr1}, \p{Tr2} are equivalent to the relation
\be
g^{-1}\left(i\epsilon_{i}Q^{i}-i\bar{\epsilon}^{i}\bar{Q}_{i}\right)g= g^{-1}\delta g + i\delta h^j_i\,I^i_j+i\delta \hat{h}\, F.
\ee
Taking into account that $g^{-1}\delta g$ is given by the same formulas \p{C1} - \p{C2}, with $\delta$ in place of $d$, it is easy to find
the basic $\epsilon$ transformations of the superspace coordinates
\bea
    &&\delta\theta_{i}=\epsilon_{i}+2m\left(\bar{\epsilon}\cdot\theta\right) \theta_{i}\,,\qquad\delta\bar{\theta}^{j}
    =\bar{\epsilon}^{i}-2m\left(\epsilon\cdot\bar\theta\,\right)\bar{\theta}^{i},\nn
    &&\delta t=i\left[\left(\epsilon\cdot\bar\theta\,\right)+\left(\bar{\epsilon}\cdot\theta\right) \right],\label{tr}
\eea
and the induced $u(2)$ elements
\bea
\delta \hat{h} &=&- {im}\left[\left(\epsilon\cdot\bar\theta\,\right)+\left(\bar{\epsilon}\cdot\theta\right) \right],\nn
   \delta h^j_i &=& {im}\left(\epsilon_i\bar{\theta}^j +\bar{\epsilon}^j \theta_i -\frac12{\delta^j_i}
    \left[\left(\epsilon\cdot\bar\theta\,\right)+\left(\bar{\epsilon}\cdot\theta\right)\right]\right)
    \left[1-\frac{m}{2}\left(\bar\theta\cdot\theta\right)\right]\nn
    &&-\,\frac{3im^2}{2}\left[\left(\epsilon\cdot\bar\theta\,\right) +\left(\bar{\epsilon}\cdot\theta\right) \right]
    \left[ \bar{\theta}^{j}\theta_{i} - \frac12{\delta^j_i}\left(\bar{\theta}\cdot\theta\right)\right] \,.\label{tr5}
\eea
The integration measure defined as
\bea
\mu :=   dt\,d^2\theta\, d^2\bar{\theta}\left[1+ 2m\left(\bar{\theta}\cdot\theta\right)\right] \label{inv}
\eea
is invariant under these transformations, $\delta \mu = 0\,$. Note that this measure can be computed by the general formula
\bea
\mu :=   dt\,d^2\theta\, d^2\bar{\theta}\;{\rm Ber}E_{M}^{\;\;\;\, A},
\eea
where ${\rm Ber}E_{M}^{\;\;\;\, A}$ is the Berezinian (superdeterminant) of the super-vielbein defined as
\bea
    E^{A}=dZ^M E_{M}^{\;\;\;\, A},\qquad dZ^M=\left(dt, d\theta_i, d\bar{\theta}^k\right),
    \qquad E^A=\left(\Delta t, \Delta \theta_i, \Delta \bar{\theta}^k\right).
\eea

{}From the general transformation law of the Cartan form $\Omega$,
\be
\Omega{\,}' = h \Omega h^{-1} -dh \,h^{-1}\,,
\ee
we find its infinitesimal transformation
\bea
    \delta\Omega =\left[\left(i\delta h^j_i\,I^i_j+i\delta \hat{h}\, F\right),\Omega\right]-dh \,h^{-1}\,.
\eea
Thus all the component Cartan forms, except those belonging to the stability subalgebra $(I^i_k, F)\,$, homogeneously transform
in $SU(2|1)\,$.

The remaining $SU(2|1)$ transformations of the superspace coordinates are contained in the closure of the $\epsilon$
and $\bar\epsilon$ transformations. They can  easily be found by computing the Lie brackets of \p{tr}.

Having found the superspace realization of  the $\epsilon$ transformations, we can define the corresponding generators
as the appropriate differential operators:
\be
    \left(\delta\theta_{i},\delta\bar{\theta}^{i},\delta t\right)
    =i\left[ \epsilon_{i}Q^{i}- \bar{\epsilon}^{j}\bar{Q}_{j}\,, \left(\theta_{i},\bar{\theta}^{i},t\right)\right],
\ee
whence
\bea
    &&Q^i=-i\frac{\partial}{\partial\theta_i}+2im\bar{\theta}^i\bar{\theta}^j\frac{\partial}{\partial\bar{\theta}^j}
    +\bar{\theta}^i\frac{\partial}{\partial t}\,,\nn
    &&\bar{Q}_j=i\frac{\partial}{\partial\bar{\theta}^j}
    +2i m\theta_j\theta_k\frac{\partial}{\partial\theta_k}-\theta_j \frac{\partial}{\partial t}\,.\label{sch}
\eea
Their anticommutators yield the superspace realization of the bosonic generators  $I^i_j, F, H$:
\bea
    &&I^i_j=\left(\bar{\theta}^i\frac{\partial}{\partial\bar{\theta}^j}-\theta_j\frac{\partial}{\partial\theta_i}\right)
    -\frac12{\delta^i_j}\left(\bar{\theta}^k\frac{\partial}{\partial\bar{\theta}^k}-\theta_k\frac{\partial}{\partial\theta_k}\right),\nn
    &&H= i\partial_{t},\qquad F=\frac{1}{2}\left(\bar{\theta}^k\frac{\partial}{\partial\bar{\theta}^k}
    -\theta_k\frac{\partial}{\partial\theta_k}\right).\label{bosch}
\eea
It is a direct exercise to check that the operators \p{sch} and \p{bosch} indeed form the $su(2|1)$ algebra \p{algebra}
with respect to (anti)commutation.

Note that the same differential operators (taken with the minus sign) realize $SU(2|1)$ on the superfields having no external $U(2)$ indices,
i.e. on the $U(2)$ scalar superfields. To construct the realization of $SU(2|1)$ on the superfields forming non-trivial $U(2)$ multiplets,
one should extend \p{sch} and \p{bosch} by the matrix parts $\delta h^j_i \tilde{I}^i_j$ and $\delta \hat{h}\tilde{F}\,$, with the parameters
$\epsilon_i\,, \,\bar\epsilon^i$ being separated. Here,  $\tilde{I}^i_j $ and $\tilde{F}$ are matrix generators of the $U(2)$ representation by which
the given superfield is rotated with respect to its external indices.

\subsection{Covariant derivatives}
The covariant derivatives of some superfield $\Phi_B(t,\theta_{i}, \bar{\theta}^{j})\,$, where $B$ is the index of some $U(2)$ representation,
can be found from the general expression for its covariant differential
\bea
{\cal D}\Phi_A := d \Phi_A + \left[i\Delta h^j_i\,\tilde{I}^i_j+i\Delta \hat{h}\,\tilde{F}\right]_A^B\Phi_B  \equiv
\left[\Delta\theta_{i} {\cal D}^{i} -\Delta\bar{\theta}^{j}\bar{{\cal D}}_{j}+ \Delta t\, {\cal D}_t\right]\Phi_A\,.
\eea
The covariant derivatives ${\cal D}^i, \bar{\cal D}_j, {\cal D}_H$ are read off from this definition as
\bea
    {\cal D}^i&=&\left[1+{m}\left(\bar{\theta}\cdot\theta\right)
    -\frac{3m^2}{4} \left(\bar{\theta}\cdot\theta\right)^2\right]\frac{\partial}{\partial\theta_i}
    - {m}\bar{\theta}^i\theta_j\frac{\partial}{\partial\theta_j}-i\bar{\theta}^i\frac{\partial}{\partial t}\nn
    &&+\, {m}\bar{\theta}^i \tilde{F}- {m}\bar{\theta}^j\tilde{I}^i_j
    +\frac{m^2}{2}\left(\bar{\theta}\cdot\theta\right)\bar{\theta}^j\tilde{I}^i_j
    -\frac{m^2}{2}\bar{\theta}^i\bar{\theta}^j\theta_k \tilde{I}^k_j\,,\nn
    \bar{{\cal D}}_j &=& -\left[1+ {m}\left(\bar{\theta}\cdot\theta\right)
    -\frac{3m^2}{4} \left(\bar{\theta}\cdot\theta\right)^2\right]\frac{\partial}{\partial\bar{\theta}^j}
    + {m}\bar{\theta}^k\theta_j\frac{\partial}{\partial\bar{\theta}^k}+i\theta_j\frac{\partial}{\partial t}\nn
    &&-\, {m}\theta_j\tilde{F}+ {m}\theta_k\tilde{I}^k_j -\frac{m^2}{2}\left(\bar{\theta}\cdot\theta\right)\theta_k\tilde{I}^k_j
    +\frac{m^2}{2}\theta_j\bar{\theta}^l\theta_k\tilde{I}^k_l\,,\nn
    {\cal D}_t &=& \partial_t\,.\label{cov}
\eea
They form the algebra which mimics $su(2|1)\,$\footnote{When computing the anticommutator of the covariant spinor derivatives,
it is assumed that the $U(2)$ matrix parts of the spinor derivative
standing on the left properly
act also on the doublet index of the derivative on the right.}
:
\bea
    &&\lbrace {\cal D}^i, \bar{{\cal D}}_j\rbrace = 2m\left( \tilde{I}^i_j -\delta^i_j \tilde{F}\right)+2i\delta^i_j{\cal D}_t \,,\qquad
    \left[\tilde{I}^i_j, \tilde{I}^k_l\right] = \delta^i_l \tilde{I}^k_j -\delta^k_j \tilde{I}^i_l  \,,\nn
    &&\left[\tilde{I}^i_j, \bar{{\cal D}}_{l}\right] = \delta^i_l\bar{{\cal D}}_{j}-\frac{1}{2}\delta^i_j\bar{{\cal D}}_{l}\, ,\qquad
    \left[\tilde{I}^i_j, {\cal D}^{k}\right] =  \frac{1}{2}\delta^i_j {\cal D}^{k} -\delta^k_j {\cal D}^{i} ,\nn
    &&\left[\tilde{F}, \bar{{\cal D}}_{l}\right]=\frac{1}{2}\bar{{\cal D}}_{l}\,,\qquad \left[\tilde{F}, {\cal D}^{k}\right]=-\frac{1}{2}{\cal D}^{k}.
\eea

\section{The  multiplet (1,4,3)}
\subsection{Constraints}
Now we are ready to define the properly constrained $SU(2|1)$ superfields encompassing the appropriate analogs of the irreducible off-shell
multiplets of the standard ${\cal N}=4, d=1$ supersymmetry. As the first example we consider an analog of the multiplet $({\bf 1, 4, 3})$
\cite{ikrlev,ikrpash}.

This multiplet is described by the real neutral superfield $G$ satisfying the $SU(2|1)$ covariantization of the standard  $({\bf 1, 4, 3})$
multiplet constraints
\bea
    \varepsilon^{lj}{\cal \bar D}_l\, {\cal \bar D}_j G =\varepsilon_{lj}{\cal D}^l\, {\cal D}^j G =0\,. \lb{143constr}
\eea
They are solved by\footnote{The superfield $G$ is fixed by \p{143constr} up to an additive constant which we set equal to zero
without loss of generality.}
\bea
    G&=&x- {mx}\left(\bar{\theta}\cdot\theta\right)[ 1 - 2m\left(\bar{\theta}\cdot\theta\right)]
    +\frac{\ddot{x}}{2}\left(\bar{\theta}\cdot\theta\right)^2 - i\left(\bar{\theta}\cdot\theta\right)\left(\theta_i \,\dot{\psi}^i
    +\bar{\theta}^j\,\dot{\bar{\psi}}_j\right)\nn
    &&+\left[1- 2m\left(\bar{\theta}\cdot\theta\right)\right]\left(\theta_i \,\psi^i-\bar{\theta}^j\,\bar{\psi}_j\right)
    + \bar{\theta}^j\theta_i \,B^i_j \,, \quad B^k_k = 0\,.\lb{Gstruct}
\eea
We see that the irreducible set of the off-shell component fields is $x(t), \psi^i(t), \bar\psi_i(t), B^i_j (t) (B^k_k =0)$, i.e., $G$ reveals
just the $({\bf 1, 4, 3})\,$ content.
In the contraction limit $m=0\,$, it is reduced to the ordinary $({\bf 1, 4, 3})$ superfield.

The $\epsilon$ transformation law of $G$,
\bea
    \delta G =-i\left[\epsilon_{i}Q^{i}- \bar{\epsilon}^{j}\bar{Q}_{j}\,, G\right],
\eea
implies the following transformation laws for the component fields:
\bea
&&\delta x =\left(\bar{\epsilon} \cdot \bar{\psi}\right)-\left(\epsilon\cdot\psi\right) ,\qquad
    \delta \psi^i =i\bar{\epsilon}^i \dot x - {m}\bar{\epsilon}^i x + \bar{\epsilon}^k B^i_k\,,\nn
    &&\delta B^i_j = -2i\left( \epsilon_j\dot{\psi}^i +\bar{\epsilon}^i\dot{\bar{\psi}}_j
    -  \frac12 {\delta^i_j}\left[\epsilon_k\dot{\psi}^k +\bar{\epsilon}^k\dot{\bar{\psi}}_k\right]\right) \nn
    && \quad \quad +\, 2m\left(  \bar{\epsilon}^i\bar{\psi}_j- \epsilon_j\psi^i
    - \frac12{\delta^i_j}\left[\bar{\epsilon}^k\bar{\psi}_k -\epsilon_k\psi^k \right]\right).\label{transf}
\eea
\subsection{The invariant Lagrangian}
Using the definition \p{inv}, we can construct the general Lagrangian and action for the $SU(2|1)$ multiplet $({\bf 1, 4, 3})\,$ as
\bea
    {\cal L}=-\int d^2\theta\, d^2\bar{\theta}\left[1+2m\left(\bar{\theta}\cdot\theta\right)\right] \, f(G)\,, \quad S = \int dt {\cal L}\,.\label{1}
\eea
Note that the superfield Lagrangian density in \p{1} is defined up to the shift
\be
f \,\rightarrow \, f + c_1\,G + c_0\,, \lb{sfFreed}
\ee
where $c_0, c_1$ are arbitrary real constants. It follows from \p{Gstruct} that after integrating over $\theta, \bar\theta$ these additional terms
yield total derivatives and so they do not contribute to the full action $S\,$.

Doing the Berezin integral in \p{1}, we obtain the component off-shell Lagrangian
\bea
    {\cal L}&=&[\dot{x}^2 +i(\bar{\psi}_i\dot{\psi}^i -\dot{\bar{\psi}}_i\psi^i)]g(x)+ \frac12{B^i_j B^j_i}g(x)
    -B^i_j \left(\frac12{\delta^j_i}\bar{\psi}_k\psi^k -\bar{\psi}_i\psi^j\right)g'(x)\nn
    && -\frac{1}{2} \left(\bar{\psi}_i\psi^i\right)^2  g''(x)+ 2m\bar{\psi}_i\psi^i g(x)+ {m x}\bar{\psi}_i\psi^i g'(x)- {m^2 x^2} g(x)\,,
\eea
where $g := f''$ and primes mean differentiation in $x\,$, $f' = \partial_x f\,,$ etc. The absence of the explicit $f$ and $f'$ in the component
action reflects the freedom \p{sfFreed}.

As the next standard step, we eliminate the auxiliary fields $B^i_j$ by their algebraic equations of motion,
\bea
     B^i_j=\frac{g'(x)}{g(x)}\left(\frac12{\delta^i_j}\bar{\psi}_k\psi^k -\bar{\psi}_j\psi^i \right),
\eea
and rewrite the Lagrangian in terms of $x$ and $\psi^i$:
\bea
    {\cal L}&=& [\dot{x}^2 + i(\bar{\psi}_i\dot{\psi}^i -\dot{\bar{\psi}}_i\psi^i)]g(x)-\frac{1}{2}
    \left(\bar{\psi}_i\psi^i\right)^2 \left[ g''(x)-\frac{3\left(g'(x)\right)^2}{2 g(x)}\right] \nn
    &&-\,{m^2 x^2} g(x)+ 2m\bar{\psi}_i\psi^i g(x)+ {m x}\bar{\psi}_i\psi^i g'(x).\label{GL}
\eea
It is invariant, modulo a total time derivative, under the following on-shell odd transformations:
\bea
    &&\delta x = \bar{\epsilon}^k \bar{\psi}_k -\epsilon_k\psi^k \,,\nn
    &&\delta \psi^i = i\bar{\epsilon}^i \dot x - {m}\bar{\epsilon}^i x
    -\left(\bar{\epsilon}^k \bar{\psi}_k\psi^i -\frac{1}{2}\bar{\epsilon}^i \bar{\psi}_k\psi^k \right)\frac{g'(x)}{ g(x)}\,, \quad \mbox{and c.c.}\,.
\eea

The Lagrangian \p{GL} can be simplified by passing to the new bosonic field $y(x)$ with the free kinetic term. From the equality
\be
\dot x^2 g(x) = \frac12 \dot y^2
\ee
we find the equation
\be
[y'(x)]^2 = 2 g(x) \; \Rightarrow \; y' = \sqrt{2 g}\,, \lb{xy}
\ee
and \p{GL} is rewritten in the form
\bea
    {\cal L}&=&\frac12 \dot{y}^2+\frac{i}{2}\,(\bar{\zeta}_i\dot{\zeta}^i-\dot{\bar{\zeta}}_i\zeta^i)- \frac{ m^2 x^2}{2}[y'(x)]^2
    + {m}\bar{\zeta}_i\zeta^i \left[1+ \frac{x y''(x)}{y'(x)}\right] \nn
    &&-\frac{1}{2}\left(\bar{\zeta}_i\zeta^i\right)^2 \left(\frac{y'''(x)y'(x)-2[y''(x)]^2}{[y'(x)]^4}\right),
\eea
where we defined $\zeta^i = \psi^i y'(x)$. Solving \p{xy} for $x$, $x = x(y)$, and defining $V(y)=x y'(x) = x(y) \frac{1}{x'(y)} $,
we obtain\footnote{The $x$ derivative of $V(y(x))$ is represented as $V_x =V'(y) y_x$}
\bea
    {\cal L}=\frac12{\dot{y}^2} +\frac{i}{2}\,(\bar{\zeta}_i\dot{\zeta}^i-\dot{\bar{\zeta}}_i\zeta^i)- \frac{m^2}{2} V^2(y)
    + {m}\bar{\zeta}_i\zeta^i V'(y) -\frac{1}{2} \left(\bar{\zeta}_i\zeta^i\right)^2 \partial_y\left[\frac{V'(y) -1}{V(y)}\right].\qquad \lb{yL}
\eea
Thus we have finally obtained the Lagrangian involving an arbitrary function $V(y)$ (it is only required to be regular at $y=0\,$).
In the new representation the supersymmetry
transformations acquire the form
\bea
    &&\delta y =\bar{\epsilon}^k \bar{\zeta}_k- \epsilon_k\zeta^k ,\nn
    &&\delta \zeta^i = i\bar{\epsilon}^i \dot y-  {m}\bar{\epsilon}^i V(y)  -  \left(\epsilon_k\zeta^k \zeta^i
    +\bar{\epsilon}^k \bar{\zeta}_k\zeta^i -\bar{\epsilon}^i \bar{\zeta}_k\zeta^k \right)\frac{V'(y) -1}{V(y)}.\lb{tranyL}
\eea

After redefinitions
\bea
    &&\zeta^i = \frac{1}{\sqrt{2}}\left(\bar{\mu}^i-\mu^i\right),\qquad \bar{\zeta}_i = \frac{1}{\sqrt{2}}\left(\bar{\mu}_i+\mu_i\right),\nn
    &&\tilde{\epsilon}_i = \frac{1}{\sqrt{2}}\left(\bar{\epsilon}_i -\epsilon_i\right),\qquad
    \tilde{\bar{\epsilon}}^j =\frac{1}{\sqrt{2}}\left(\bar{\epsilon}^j +\epsilon^j \right),\qquad  \tilde{V}(y)= m V(y),
\eea
the Lagrangian \p{yL} and the transformation rules \p{tranyL} are recognized as defining the general SQM model with
``weak'' ${\cal N}=4$ supersymmetry \cite{WS}. Thus this model is the on-shell version of the general $SU(2|1)$ symmetric model of a single
$({\bf 1, 4, 3})$ multiplet. In what follows, we will stick to our original choice of the fermionic variables.

\setcounter{equation}{0}
\section{Quantum (1,4,3) oscillator model}
\subsection{Basics}
Let us consider the simplest  Lagrangian
\bea
    {\cal L}&=&\frac12{ \dot{x}^2}-\frac12{m^2 x^2}+\frac{i}{2}(\bar{\psi}_i \dot{\psi}^i
    -\dot{\bar{\psi}}_i \psi^i)+ {m}\,\bar{\psi}_i {\psi}^i ,\label{OL}
\eea
which corresponds to the choice
\bea
g = 1/2 \;\Rightarrow \;    f(x)=\frac14{x^2} + c_0 x + c_1\,,
\eea
in \p{GL}.
This Lagrangian is invariant under the transformations
\bea
    \delta x =\left(\bar{\epsilon} \cdot \bar{\psi}\right)-\left(\epsilon\cdot\psi\right),\qquad
    \delta \psi^i = i\bar{\epsilon}^i \dot x- {m}\bar{\epsilon}^i x\,.
\eea
The corresponding conserved Noether charges are easily calculated to be:
\bea
    &&  Q^i = \psi^i\left(p - {i m x} \right) ,\qquad
    \bar{Q}_i =\bar{\psi}_i \left( p +  {i m x} \right),\nn
    &&  F = \frac{1}{2}\psi^k \bar{\psi}_k\, , \qquad I^i_j = \psi^i\bar{\psi}_j  -\frac{1}{2}\delta_j^i\psi^k \bar{\psi}_k\,.
\eea
The Poisson brackets are imposed as\footnote{For fermionic fields these are in fact Dirac brackets.}
\bea
    \{x, p\} =1\,, \qquad \{\psi^i , \bar{\psi}_j \} = -i\delta^i_j \,.\lb{brack}
\eea
The corresponding canonical Hamiltonian reads
\bea
    H&=&\frac{p^2}{2} +\frac{m^2 x^2}{2}+ {m}\psi^i \bar{\psi}_i\,.\lb{Hcl}
\eea
Its bosonic part is just the Hamiltonian of harmonic oscillator. We quantize the brackets \p{brack} in the standard way
\be
 [ \hat x, \hat p] =i\,, \quad \{\hat{\psi}^i , \hat{\bar{\psi}}_j \} = \delta^i_j\,, \quad  \hat p = -i\partial_x\,, \;
 \hat{\bar{\psi}}_j = \partial/\partial \hat{\psi}^j\,,\lb{quantize}
\ee
and use the relation
\be
\left[\left( \hat p - {i m \hat x} \right),\left( \hat p +  {i m \hat x} \right)\right] = 2m
\ee
to represent the quantum Hamiltonian as
\bea
    \hat H&=&\frac{1}{2}\left( \hat p  +  {im \hat x} \right)\left( \hat p  - {i m \hat x} \right)
    + {m} \hat{\psi}^i\hat{\bar{\psi}}_i\,  .\label{H}
\eea

The quantum operators associated with the remaining Noether charges are
\bea
    && \hat{Q}^i = \hat{\psi}^i\left(\hat p - {im \hat x}\right) ,\qquad
    \hat{\bar{Q}}_i = \hat{\bar{\psi}}_i \left(\hat p + {i m \hat x}\right) ,\lb{hatQ} \\
    && \hat{F} = \frac{1}{2}\hat{\psi}^k \hat{\bar{\psi}}_k\, , \qquad \hat{I}^i_j =  \hat{\psi}^i \hat{\bar{\psi}}_j
    - \frac{1}{2}\delta_j^i \hat{\psi}^k \hat{\bar{\psi}}_k\,.\lb{hatFI}
\eea
One can check that they indeed form the superalgebra $su(2|1)$:
\bea
    &&\lbrace\hat{Q}^i,\hat{\bar{Q}}_j\rbrace =2\delta_j^i \hat H + 2m\left(  \hat{I}_j^i - \delta_j^i \hat{F}\right),\nn
    &&\left[\hat{I}^i_j, \hat{\bar{Q}}_{l}\right] =\frac{1}{2}\delta^i_j\hat{\bar{Q}}_{l} -\delta^i_l\hat{\bar{Q}}_{j}\, ,\qquad
    \left[\hat{I}^i_j,\hat{Q}^{k}\right] = \delta^k_j \hat{Q}^{i} - \frac{1}{2}\delta^i_j \hat{Q}^{k},\nn
    &&\left[\hat{I}^i_j,  \hat{I}^k_l\right] =  \delta^k_j \hat{I}^i_l- \delta^i_l \hat{I}^k_j \,,\qquad\left[\hat{F}, \hat{\bar{Q}}_{l}\right]
    =-\frac{1}{2}\hat{\bar{Q}}_{l}\,,\qquad \left[\hat{F}, \hat{Q}^{k}\right]=\frac{1}{2}\hat{Q}^{k}.\label{salgebra}
\eea
Note that there is a freedom of adding some constants to $\hat H$ and $\hat F$, in such a way that the sum $\hat H -m\hat{F}$ remains intact.
Using this freedom, one can, e.g., cast $\hat{H}$ in the form which corresponds to just making replacements \p{quantize} in the classical
Hamiltonian \p{Hcl}. In what follows, we will deal with the quantum operators defined as in \p{H} - \p{hatFI}.

For further use, we give the expressions for the second- and third-order  Casimir operators $C_2, C_3$ of $SU(2|1)$.
The explicit form of these operators in terms of the $SU(2|1)$ generators can be found, e.g., in \cite{ssph,04cos1}. We will use the following concise
representation for the Casimirs,
\be
4m^2C_2 = C^i_i\,, \quad 12\, m^3 C_3 = 6m^3{F}'(1 + 2C_2) + m\, I^i_k C^k_i\,,\label{eq413}
\ee
where
\be
{F}' = F -\frac{1}{m}\,H\,, \quad C^i_j = 2{m^2}\,\delta^i_j \,{F'}^2
- {m^2}\left\lbrace I_l^i, I_j^l \right\rbrace + {m}\left[Q^i, \bar{Q}_j\right].
\ee

These expressions are valid irrespective of the particular realization of the $SU(2|1)$ generators. For our quantum-mechanical realization
\p{H} - \p{hatFI}
they are reduced to the following nice form
\bea
    m^2 C_2 = \hat H\left(\hat H-m\right),\qquad
    m^3C_3  = \hat{H} \left(\hat H-m\right)\left(\hat H-\frac{m}{2}\right). \lb{Casimhat}
\eea
Thus they are fully specified by the energy spectrum of the quantum Hamiltonian.

\subsection{Wave functions and spectrum}

We construct the Hilbert space of wave functions  in terms of wave functions of bosonic harmonic oscillator, to which the system \p{hatQ}, \p{H}
is reduced, when discarding the fermions.

The generic super wave function $\Omega^{(\ell)}$ at the energy level $\ell$ shows up the four-fold degeneracy due
to the $\psi$-expansion\footnote{Our definition of super wave functions  is different from that of \cite{WS} because of a different choice of the
fermionic variables. The two sets are related through the similarity transformation $\psi \rightarrow \chi\,, \Omega \rightarrow \tilde{\Omega}\,$, with
$$
    \chi^k =-e^{-U} \psi^k\, e^{U}\,,\quad \bar{\chi}_k= e^{-U} \bar{\psi}_k \,e^{U}\,,\quad \tilde{\Omega}=e^{-U} \Omega \,,\quad
    U =\frac{\pi}{8}\,(\varepsilon_{ij}\psi^i\psi^j -\varepsilon^{ij}\bar{\psi}_i\bar{\psi}_j)\,.
$$}
\bea
    \Omega^{(\ell )} = a^{(\ell )}|\ell\rangle + b^{(\ell )}_i\,\psi^i  |\ell -1\rangle + \frac{1}{2}\,c^{(\ell )}\,
    \varepsilon_{ij} \psi^i \psi^j |\ell-2\rangle ,
    \qquad  \ell \geq 2\, ,\lb{ell2}
\eea
where $|\ell\rangle , |\ell -1\rangle , |\ell -2\rangle$  are the harmonic oscillator functions at the relevant levels and
$a^{(\ell)}, b^{(\ell)}, c^{(\ell)}$
are some numerical coefficients. We treat the operators $\hat p \pm im {x}$ in \p{H} and \p{hatQ} as the creation
and annihilation operators and impose
the standard physical conditions
\bea
    \hat{\bar{\psi}}_k|\ell\rangle =0,\qquad \left( \hat p - {i m \hat x} \right)|0\rangle =0,\qquad
    \left( \hat p +  {i m \hat x} \right)|\ell\rangle
    =|\ell +1\rangle  .
\eea
The spectrum of the Hamiltonian \p{H} is then
\bea
    \hat H \,\Omega^{(\ell)}= {m}\,\ell\,\Omega^{(\ell)},\qquad m>0,\quad \ell \geq 0.\lb{spectr}
\eea

We observe that the ground state ($\ell =0$) and the first excited states ($\ell =1$) are special, in the sense that they encompass
non-equal numbers
of bosonic and fermionic states:
\bea
\Omega^{(0)} = a^{(0)}\,  | 0\rangle\,, \qquad     \Omega^{(1)} = a^{(1)}\,  | 1\rangle +b^{(1)}_i \,\psi^i  |0\rangle\,.\lb{ell01}
\eea

The ground state is annihilated by all $SU(2|1)$ generators including $Q^i$ and $\bar Q_i$, so it is an $SU(2|1)$ singlet. The states with $\ell =1$
can be shown to form the fundamental representation of $SU(2|1)$. The action of the supercharges on these states is given by
\bea
&& Q^i\, \psi^k| 0\rangle =0\,, \qquad \bar Q_i\, \psi^k| 0\rangle = \delta^k_i\, | 1\rangle\,, \nn
&& Q^i\, | 1\rangle = 2m\,\psi^i| 0\rangle\,, \qquad \bar Q_i\, | 1\rangle =0\,. \lb{fundQQ}
\eea

It is instructive to see what values the Casimir operators \p{Casimhat} take on all these
states.
The values of Casimir operators for the finite-dimensional $SU(2|1)$ representations can be written in the following generic form \cite{repres}
\be
    C_2 = (\beta^2 - \lambda^2), \quad C_3 = \beta(\beta^2 - \lambda^2) = \beta C_2. \lb{betalambda}
\ee
These representations are characterized by some positive number $\lambda$ (``highest weight''), which can be half-integer or integer,
and an arbitrary additional real number $\beta\,$,  which is related to the eigenvalues of the internal $U(1)$ generator ${F}'\,$.
Comparing \p{betalambda} with the expressions \p{Casimhat} and using the formula for the energy spectrum  \p{spectr},
we find that in our case $\lambda =1/2$ for any $\Omega^{(\ell)}$ and
\be
    C_2(\ell) = \left(\ell - 1 \right)\ell, \quad C_3(\ell) = \left(\ell  - 1/2 \right)\left(\ell - 1 \right)\ell ,\quad \beta(\ell) =
    \ell  - 1/2\,. \label{Casimnumbers}
\ee
The ground state with $\ell=0$ is atypical, because Casimir operators take zero values on it.
On the states with $\ell =1$ both Casimirs  vanish as well, so these states also form an atypical $SU(2|1)$ representation.
On the $\ell >1$ states both Casimirs are non-zero,
so these states belong to the typical $SU(2|1)$ representations characterized by equal numbers of the bosonic and fermionic states.

Defining the inner product of the states as
\be
    \langle \Omega |\Psi\rangle = \int_{-\infty}^{\infty} dx\,\Omega^\dagger \Psi\,,
\ee
one can check that the states $\Omega^{(\ell)}$ for different $\ell$ are orthogonal with respect to this product and the norms of these states
are positive-definite. For instance,
\bea
\langle 0|0 \rangle =\int_{-\infty}^{\infty} dx \,\exp{\left(-m x^2\right)} = \sqrt{\frac{\pi}{m}}\,.
\eea
The norm of the state $\Omega^{(\ell)}$ is defined as
\be
    ||\Omega^{(\ell)}||^2 = \frac{\langle \Omega^{(\ell)}|\Omega^{(\ell)} \rangle}{ \langle \ell|\ell \rangle}.
\ee
Hence, for the wave functions \p{ell2}, \p{ell01} we find the following manifestly positive norms:
\bea
    &&||\Omega^{(\ell)}||^2 = \bar{a}^{(\ell)}a^{(\ell)}+ \frac{\bar{b}^{(\ell)i}b^{(\ell)}_i} {2m \ell}
    + \frac{\bar{c}^{(\ell)}c^{(\ell)}}{4m^2\left(\ell -1\right)\ell},\quad \ell \geq 2 ,\nn
    &&||\Omega^{(1)}||^2 = \bar{a}^{(1)} a^{(1)}+\frac{\bar{b}^{(1)i} b^{(1)}_i}{2m},\quad  ||\Omega^{(0)}||^2 = \bar{a}^{(0)} a^{(0)}.\label{anorm}
\eea
\subsection{Exotic $SU(2)$ symmetry}
Let us define the operators $\hat{B}_\pm$ belonging to the universal enveloping algebra of $su(2|1)$:
\bea
    \hat{B}_+= \hat{\bar{Q}}_2\hat{\bar{Q}}_1=\left(\hat p + {i m \hat x}\right)^2 \hat{\bar{\psi}}_2\hat{\bar{\psi}}_1 \,, \qquad \hat{B}_-
    = \hat{Q}^1\hat{Q}^2=\left(\hat p - {i m \hat x}\right)^2\hat{\psi}^1\hat{\psi}^2\,.\lb{Bpm}
\eea
Defining also the operator
\be
\hat{B}_3=2 m^2 C_2\left(1-\hat{\psi}^k \hat{\bar{\psi}}_k\right),\lb{B3}
\ee
we observe that at every level with $C_2\neq 0$, i.e. with $\ell \geq 2$, these three operators generate the algebra $su(2)$:
\bea
\left[\hat{B}_+ ,\hat{B}_-\right]=2\hat{B}_3\,,\quad \left[\hat{B}_3 ,\hat{B}_\pm\right]=\pm 4 m^2 C_2\hat{B}_\pm\,.
\eea
The whole algebra is non-vanishing only on the bosonic states ($|\ell\rangle\,, \;\varepsilon_{ij} \psi^i \psi^j  |\ell-2 \rangle$), which form
doublets of this $su(2)$. The fermionic states are singlets. This  extra $su(2)$ algebra commutes with the Hamiltonian and with the
$su(2)$ subalgebra of $su(2|1)$, while the fermionic $U(1)$ charge operator $\hat{F}$ defines its outer automorphism.
It is instructive to quote the Casimir operator of this  $su(2)$ algebra
\bea
    B^2 =2 m^2 C_2\left\{B_+ ,B_-\right\}+(B_3)^2 = 12\, m^4 (C_2)^2\left(1-\hat{\psi}^k \hat{\bar{\psi}}_k\right)^2.
\eea

Using the definition \p{Bpm}, it is also easy to show that
\bea
    \left\{\hat{B}_+ ,\hat{B}_-\right\}= 4\, m^2 C_2\left(1-\hat{\psi}^k \hat{\bar{\psi}}_k\right)^2.
\eea
On the fermionic states $\psi^i  |\ell -1\rangle $ this anticommutator  vanishes, while on the bosonic states it becomes
\bea
    \left\{\hat{B}_+ ,\hat{B}_-\right\}= 4\, m^2 C_2  =4 \hat{H}(\hat{H} - m)\,.\lb{2fold}
\eea
Thus, the operators $\hat{B}_+ ,\hat{B}_-$ can be interpreted as generators of some ``${\cal N}=2$ superalgebra'' acting only on the bosonic states
and possessing the ``Hamiltonian'' which is quadratic in the Hamiltonian of the original $SU(2|1)$ invariant system. Just this ``superalgebra''
was constructed in \cite{WS} in order to establish a link with the so called ``$N$-fold'' supersymmetries, which are defined
by the nonlinear algebras of the type \p{2fold} (see \cite{Nfold} and references therein). Our consideration in the framework
of the simple oscillator model shows that the relevant product
``supercharges'' \p{Bpm} are in fact generators of some extra bosonic $su(2)$ algebra which belongs to the universal enveloping of $su(2|1)$ and is
such that the full space of quantum states is split into its doublets and singlets. The relevant nonlinear ``Hamiltonian'' proves to be the quadratic
Casimir of $su(2|1)$. It is an open question whether this interpretation applies to the case of the general quantum $({\bf 1, 4, 3})$ models.

\setcounter{equation}{0}
\section{The multiplets (2,4,2)}
\subsection{Chiral $SU(2|1)$ superspaces}
The supergroup $SU(2|1)$ admits two mutually conjugated complex supercosets which can be identified with the left and right chiral subspaces:
\bea
    \left(t_L,\theta_i\right),\qquad  \left(t_R, \bar{\theta}^i\,\right). \lb{LeftRight}
\eea
The relevant complex even coordinates are related to the real time coordinate $t$ via
\bea
    t_L =t+\frac{i}{2m}\ln{K},\quad t_R =t-\frac{i}{2m}\ln{K},\qquad
    K=\left[1+ 2 m\left(\bar{\theta}\cdot\theta\right)\right].\lb{tLtRt}
\eea
The Grassmann coordinates $\theta_i$ and $\bar\theta^i$ are same as in \p{tr}. The relations \p{tLtRt} uniquely follow, up to unessential
shifts $t_L \rightarrow t_L + a\theta^2\,, t_R \rightarrow t_R + \bar a\bar\theta^2\,,$ from requiring the sets \p{LeftRight} to be
closed under the $SU(2|1)$ transformations.  The latter act on the so defined  coordinates $\left(t_L,\theta_i\right),$
$\left(t_R, \bar{\theta}^i\,\right)$
as
\bea
    \delta\theta_{i}=\epsilon_{i}+2m\left(\bar{\epsilon}\cdot\theta\right) \theta_{i}\,,
    \qquad \delta t_L=2i\left(\bar{\epsilon}\cdot\theta\right), \quad
    \mbox{and c.c.}\,.
\eea

The multiplet $({\bf 2,4,2})$ is described by a  complex superfield $\Phi$ subjected to the chirality condition and possessing
a fixed external $U(1)$ charge\footnote{In principle, we could ascribe to it also a non-trivial external $SU(2)$ index, but we do not consider here
such complications.}
\bea
    \bar{{\cal D}}_j\Phi_L = 0 , \qquad \tilde{I}_j^i\Phi =0,\qquad \tilde{F}\Phi =2\kappa \Phi . \lb{chircond}
\eea
The general solution of \p{chircond} reads:
\bea
    &&\Phi(t,\theta,\bar{\theta}\,) = e^{2 i\kappa m \left(t_L - t\right)}\Phi_L (t_L,\theta)
    =\left[1+ 2m\left(\bar{\theta}\cdot\theta\right) \right]^{-\kappa}\Phi_L (t_L,\theta),\nn
    &&\Phi_L (t_L,\theta)= z+\sqrt{2}\,\theta_i \xi^i +\varepsilon^{ij}\theta_i\theta_j B .\label{chiral}
\eea
In the central basis $\{t, \theta_i, \bar\theta^k\}$ the same superfield is written as
\bea
    \Phi(t,\theta,\bar{\theta}\,)&=& z+\sqrt{2}\,\theta_i \xi^i +\varepsilon^{ij}\theta_i\theta_j B+ i\left(\bar{\theta}\cdot\theta\right)\nabla_t z
    +\sqrt{2}\,i\left(\bar{\theta}\cdot\theta\right)\theta_i \nabla_t{\xi}^i \nn
    &&-\frac{1}{2} \left(\bar{\theta}\cdot\theta\right)^2 \left[2 i m\nabla_t z+\nabla_t^2 z\right]     ,
\eea
where
\bea
    \nabla_t = \partial_t +2 i\kappa m, \qquad \bar{\nabla}_t = \partial_t - 2 i\kappa m\,.
\eea

The chiral superfield having the $\tilde{F}$ charge $2\kappa$ transforms as
\bea
&&\delta\Phi \simeq \Phi'(t',\theta',\bar\theta') - \Phi (t, \theta, \bar\theta) =\left(i\delta \hat{h}\tilde{F}
+i\delta h^j_i\tilde{I}_j^i\right)\Phi =2\kappa m\left(\epsilon_{i}\bar{\theta}^{i} +\bar{\epsilon}^{j}\theta_{j}\right)\Phi\; \;
\Longleftrightarrow \nn
    &&\delta\Phi_L (t_L,\theta) =4\kappa m\left(\bar{\epsilon}^{j}\theta_{j}\right)\Phi_L (t_L,\theta). \label{passive}
\eea
The corresponding generators $Q^i, \bar Q_i$ can be easily found, but we do not quote them here.

The superfield transformation laws \p{passive} induce the following transformations for the component fields
\bea
    &&\delta z  =-\sqrt{2}\,\epsilon_i \xi^i ,\qquad\delta \xi^i =  \sqrt{2}\, i\bar{\epsilon}^i\nabla_t z
    -\sqrt{2}\,\varepsilon^{ik}\epsilon_k B ,\nn
    &&\delta B = -\sqrt{2}\,\varepsilon_{ik}\bar{\epsilon}^k \left[m \xi^i + i\nabla_t{\xi}^i\right].\lb{susyoff}
\eea
\subsection{The Lagrangian}
The general Lagrangian involves the function $ f\left(\Phi ,\Phi^\dagger\right)$, which is an analog of the  K\"ahler potential of the
standard ${\cal N}=4$ mechanics based on the multiplet $({\bf 2, 4, 2})$ \cite{242stand}:
\bea
    {\cal L}_{(k)}=\frac{1}{4}\int d^2\theta\, d^2\bar{\theta}\left[1+2m\left(\bar{\theta}\cdot\theta\right)\right] f\left(\Phi , \Phi^\dagger\right).
    \label{kinterm}
\eea

If $\kappa =0$, the potential $ f\left(\Phi ,\Phi^\dagger\right)$
can be an arbitrary function of its arguments, without breaking of the $U(1)$ invariance generated by $F$. For $\kappa\neq 0$,
the $U(1)$ invariance necessarily implies that $ f\left(\Phi,\Phi^\dagger\right) = \tilde{f}\left(\Phi \Phi^\dagger\right)$
because the $U(1)$ generator
$F$ has a non-trivial matrix part in this case, $\delta\Phi = 2i\kappa \alpha \Phi\,,\; \delta\bar\Phi = -2i\kappa \alpha \bar\Phi\,$.

The general component Lagrangian reads:
\bea
    {\cal L} &=& -\frac{1}{4} f_z\nabla_t^2 z -\frac{1}{4} f_{\bar{z}}\bar{\nabla}_t^2{\bar{z}}-\frac{1}{4}f_{z z}\left(\nabla_t z\right)^2
    - \frac{1}{4}f_{\bar{z}\bar{z}}\left(\bar{\nabla}_t{\bar{z}}\right)^2+ \frac{1}{2}g\bar{\nabla}_t {\bar{z}}\nabla_t {z}\nn
    &&    -\,\frac{i m}{2}\left(f_{\bar{z}}\bar{\nabla}_t{\bar{z}} -f_z\nabla_t{z}\right)-\frac{i}{2}\left(\bar{\xi}\cdot\xi\right)
    \left(\bar{\nabla}_t{\bar{z}}g_{\bar{z}} -\nabla_t{z}g_z\right)
    +\frac{i}{2}\left(\bar{\xi}_i \nabla_t{\xi}^i+ \xi^i \bar{\nabla}_t{\bar{\xi}}_i\right)g\nn
    &&+\, m \left(\bar{\xi}\cdot\xi\right)g
    + \frac{1}{2}\left(\bar{\xi}\cdot\xi\right)^2 g_{z\bar{z}}+B\bar{B}g+\frac{1}{2}\varepsilon_{kl}\xi^k\xi^l
    \bar{B}g_z+\frac{1}{2}\varepsilon^{kl}\bar{\xi}_k\bar{\xi}_l B g_{\bar{z}}\,,\lb{Loff}
\eea
where $g :=f_{z \bar{z}}$ is the metric on a K\"ahler manifold\footnote{Here, the lower case indices denote the differentiation
in $z,\bar{z}$:   $f_{z \bar{z}}=\partial_z \partial_{\bar z}f$.}. After eliminating the auxiliary field $B$ by its equation of motion,
\bea
    &&B =- \frac{1}{2 g}\varepsilon_{kl}\xi^k\xi^l  g_z\,,
\eea
the Lagrangian can be rewritten as
\bea
    {\cal L} &=& g\dot{\bar{z}}\dot{z} +2 i\kappa m \left(\dot{\bar{z}}z -\dot{z}\bar{z}\right)g -\frac{i m}{2}\left(\dot{\bar{z}}f_{\bar{z}}
    -\dot{z}f_z\right)
    -\frac{i}{2}\left(\bar{\xi}\cdot\xi\right) \left(\dot{\bar z}g_{\bar{z}}-\dot{z}g_z\right)
    \nn
    &&+ \frac{i }{2}\left(\bar{\xi}_i\dot{\xi^i}-\dot{\bar{\xi}}_i\xi^i \right)g - m^2 V  -m\left(\bar{\xi}\cdot\xi\right)U
    + \frac{1}{2}\left(\bar{\xi}\cdot\xi\right)^2 R ,\lb{Lon}
\eea
where
\bea
    V &=& \kappa \left(\bar{z}\partial_{\bar{z}} +z\partial_z \right)f-\kappa^2 \left(\bar{z}\partial_{\bar{z}} +z\partial_z \right)^2 f, \nn
    U &=&\kappa \left(\bar{z}\partial_{\bar{z}} +z\partial_z \right)g -\left(1-2\kappa\right)g\,,\nn
    R &=&g_{z\bar{z}} -\frac{g_z g_{\bar{z}}}{g} \,.
\eea
  The on-shell transformations read
\bea
    &&\delta z  = -\sqrt{2}\,\epsilon_i \xi^i ,\qquad\delta \xi^i =  \sqrt{2}\,i\bar{\epsilon}^i\nabla_t z
    +\sqrt{2}\, \epsilon_k \xi^k \xi^i  \frac{g_z}{g}.
    \lb{susyon}
\eea
It is worth pointing out that, at $\kappa \neq 0\,$, one {\it has} to choose $f(z, \bar z) = \tilde{f}(z\bar z)\,$ in both the off-shell
and the on-shell component
Lagrangians \p{Loff} and \p{Lon}. Only under this restriction the $\kappa \neq 0\,$ Lagrangians are invariant, modulo a total derivative,
with respect to
the transformations \p{susyoff} and \p{susyon}.

To close this subsection, let us summarize a few peculiar features of the Lagrangian \p{Lon} at $m\neq 0$, which distinguish it from
its standard K\"ahler $({\bf 2, 4, 2})$ counterpart (recovered in the limit $m=0\,$).
\begin{itemize}
\item The $m\neq 0$ Lagrangian contains the bosonic potential $V(z, \bar z)$ which is expressed in terms of the ``K\"ahler potential'' $f$
and vanishes
at $\kappa = 0$.

\item In addition, there is a new Yukawa-type coupling $\sim U$ which is also determined by $f$ and survives at $\kappa = 0\,$.

\item The $m\neq 0$ Lagrangian contains two  $d=1$ WZ terms $\sim \kappa m$ and $\sim m$. At $\kappa = 0$, one of them vanishes,
while the other retains.

\item These WZ terms necessarily accompany the K\"ahler kinetic term $\sim \dot z\dot{\bar{z}}$ and so are prescribed by the $SU(2|1)$ supersymmetry.
No such terms can be defined for the standard linear ${\cal N}=4, d=1$ chiral multiplet $({\bf 2, 4, 2})\,$ \cite{242stand}.
\end{itemize}

\subsection{Superpotential}
When $\kappa\neq 0$, we can also add to ${\cal L}_{(k)}$ the potential term
\bea
    {\cal L}_{(p)} &=& \tilde{m}\left[\int d^2\theta \,{\cal F}\left(\Phi_L \right)+ \mbox{c.c.}\right].\lb{poten1}
\eea
As opposed to the case of standard ${\cal N}=4$ mechanics \cite{242stand}, in the $SU(2|1)$ case the superfield potential
${\cal F}$ is severely constrained by the requirement of compensating the non-trivial transformation of the chiral measure $dt_Ld^2\theta$:
\bea
    \delta dt_Ld^2\theta = -2m\, dt_L d^2\theta \,(\bar{\epsilon}^{j}\theta_{j}).
\eea
The only possibility to ensure the invariance is to choose the potential as
\bea
    {\cal L}_{(p)} &=& \tilde{m}\left[\int d^2\theta \left(\Phi_L \right)^{{1\over 2\kappa}}+ \mbox{c.c.}\right] =
    {\tilde{m} \over \kappa}\left[B z^{({1\over 2\kappa}-1)}+\bar{B} \bar{z}^{({1\over 2\kappa}-1)}\right] \nn
   && +\,{\tilde{m} \over 2\kappa}\left({1\over 2\kappa}-1\right)\left[\varepsilon_{ik}\xi^i\xi^k z^{({1\over 2\kappa}-2)}+ \varepsilon^{ik}
   \bar{\xi}_i\bar{\xi}_k \bar{z}^{({1\over 2\kappa}-2)}\right],\lb{poten2}
\eea
where $\tilde{m}$ is an extra parameter of the mass dimension. The potential term takes the simplest form $\sim B + \bar{B}$ at $2\kappa =1$.
For $\kappa =0\,$, no potential terms are possible at all. For simplicity, in what follows we will
limit our consideration to the option $\tilde{m}=0\,$.

\subsection{Hamiltonian formalism}
Performing the Legendre transformation, we define the classical Hamiltonian as:
\bea
    H &=&g^{-1}\left( p_z -\frac{i }{2}m f_z +\frac{i}{2}g_z\,\xi^k\bar{\xi}_k+2 i\kappa m \bar{z}g \right)
    \left(p_{\bar{z}}+\frac{i }{2}m f_{\bar{z}} -\frac{i}{2}g_{\bar{z}}\,\xi^k\bar{\xi}_k-2 i\kappa m z g\right)\nn
    &&+  m^2 V +m\left(\bar{\xi}\cdot\xi\right)U -\frac{1}{2}\left(\bar{\xi}\cdot\xi\right)^2 R\, .\lb{Ham242}
\eea
By Noether prescription we can calculate the supercharges $\left(Q^i\right)^\dagger =\bar{Q}_i$ and the remaining bosonic charges:
\bea
    &&Q^i =\sqrt{2}\,\xi^i\left( p_z -\frac{i }{2}m f_z +\frac{i}{2}g_z\,\xi^k\bar{\xi}_k\right),\qquad
    \bar{Q}_j = \sqrt{2}\,\bar{\xi}_j\left(p_{\bar{z}}+\frac{i }{2}m f_{\bar{z}} -\frac{i}{2}g_{\bar{z}}\, \xi^k\bar{\xi}_k\right) ,\nn
    &&F =- 2i\kappa\left(z p_z -\bar{z}p_{\bar{z}}\right) -\left(2\kappa-\frac{1}{2}\right)g\,\xi^k \bar{\xi}_k\, ,
    \qquad I^i_j =g\left[\xi^i\bar{\xi}_j  -\frac{1}{2}\delta_j^i\xi^k \bar{\xi}_k \right].\label{NC}
\eea
{}For $\kappa\neq 0\,$,
\bea
f\left(\Phi ,\Phi^\dagger\right) = \tilde{f}\left(\Phi \Phi^\dagger\right) \Rightarrow f(z,\bar{z})= \tilde{f}(z\bar{z}),
\eea
and we can cast the Hamiltonian \p{Ham242} in the following form
\bea
    H &=&g^{-1}\left( p_z -\frac{i }{2}m f_z +\frac{i}{2}g_z\xi^k\bar{\xi}_k\right)
    \left(p_{\bar{z}}+\frac{i }{2}m f_{\bar{z}} -\frac{i}{2}g_{\bar{z}}\xi^k\bar{\xi}_k\right)- 2i\kappa m\left(z p_z -\bar{z}p_{\bar{z}}\right) \nn
    &&-m\left(1-2\kappa\right)\left(\bar{\xi}\cdot\xi\right)g -\frac{1}{2}\left(\bar{\xi}\cdot\xi\right)^2 R .\label{Hcl242}
\eea
The $\kappa = 0$ form of the Hamiltonian \p{Ham242} coincides with that of \p{Hcl242} without any restrictions on the function $f(z, \bar z)$.
One of the admissible choices of $f(z, \bar z)$ in the $\kappa = 0$ case is, as before, $f(z, \bar z) = \tilde{f}(z\bar z)\,$.

The Poisson (Dirac) brackets are imposed as:
\bea
    \{z, p_z\} =1, \qquad \{\xi^i ,\bar{\xi}_j \} = -i\delta^i_j \, g^{-1}\,.
\eea
To prepare the system for quantization, it is useful to make the substitution
\be
    \left(z, \xi^i\right)\longrightarrow \left(z, \eta^i\right),\qquad\eta^i = g^{{1\over 2}}\xi^i. \label{eta-xi}
\ee
In terms of the new variables, the brackets become
\bea
    \{z, p_z\} =1, \qquad \{\eta^i ,\bar{\eta}_j \} = -i\delta^i_j \,,\qquad \{p_z , \eta^i\} = \{p_z , \bar{\eta}_j \}=0\,.\lb{brack1}
\eea
The Noether charges \p{NC} and the Hamiltonian \p{Hcl} are rewritten as
\bea
    Q^i &=&\sqrt{2}\,\eta^i g^{-{1\over 2}}\left( p_z -\frac{i }{2}m f_z +\frac{i}{2}g^{-1}g_z \,\eta^k\bar{\eta}_k\right),\nn
    \bar{Q}_j &=& \sqrt{2}\,\bar{\eta}_j\,g^{-{1\over 2}}\left(p_{\bar{z}}+\frac{i }{2}m f_{\bar{z}}
    -\frac{i}{2}g^{-1}g_{\bar{z}}\,\eta^k\bar{\eta}_k\right),
    \lb{Supclass1} \\
    F &=& - 2i\kappa\left(z p_z -\bar{z}p_{\bar{z}}\right) -\left(2\kappa-\frac{1}{2}\right)\eta^k \bar{\eta}_k , \qquad
    I^i_j =\eta^i\bar{\eta}_j  -\frac{1}{2}\delta_j^i\eta^k \bar{\eta}_k \,,\label{NC1} \\
    H &=&g^{-1}\left( p_z -\frac{i }{2}m f_z +\frac{i}{2}g^{-1}g_z\, \eta^k\bar{\eta}_k\right)
    \left(p_{\bar{z}}+\frac{i }{2}m f_{\bar{z}} -\frac{i}{2}g^{-1}g_{\bar{z}}\,\eta^k\bar{\eta}_k\right) \nn
    &&- 2i\kappa m\left(z p_z -\bar{z}p_{\bar{z}}\right)+m\left(1-2\kappa\right)\eta^k\bar{\eta}_k -\frac{1}{2}\,
    g^{-2}R\left(\eta^k\bar{\eta}_k\right)^2 .\label{Hcl1}
\eea
\subsection{Quantization}
We quantize the brackets \p{brack1} in the standard way,
\bea
    &&[ \hat z, \hat{p}_z] =i\,, \qquad \{\hat{\eta}^i , \hat{\bar{\eta}}_j \} = \delta^i_j\,,\qquad [\hat{p}_z , \hat{\eta}^i ]
    = [\hat{p}_z , \hat{\bar{\eta}}_j ]=0\,,
    \nn &&  \hat p_z = -i\partial_z\,, \qquad \hat{\bar{\eta}}_j = \frac{\partial}{\partial \hat{\eta}^j}\,,
\eea
and use the relation
\bea
    \left[\nabla_z ,\bar{\nabla}_{\bar{z}}\right] = m g -\frac{1}{2}\,g^{-1} R\left(\hat{\eta}^k\hat{\bar{\eta}}_k
    -\hat{\bar{\eta}}_k\hat{\eta}^k\right) ,
    \eea
where
\bea
    \nabla_z &= & -i\partial_z  -\frac{i }{2}m f_z +\frac{i}{2}g^{-1}g_z\left(\hat{\eta}^k\hat{\bar{\eta}}_k- 1\right),\nn
    \bar{\nabla}_{\bar{z}}&=&-i\partial_{\bar{z}}+\frac{i }{2}m f_{\bar{z}}
    +\frac{i}{2}g^{-1}g_{\bar{z}}\left(\hat{\bar{\eta}}_k\hat{\eta}^k - 1\right).
\eea

The general scheme of passing from the classical supercharges to the quantum ones was described in \cite{How}. It involves two steps.
\begin{enumerate}
\item First, one has to  Weyl-order the supercharges. The Weyl-ordered supercharges act on super wave functions with the inner product
\bea
    \langle \Omega |\Psi\rangle =\int dz\,d\bar{z}\prod_i d\eta^i\,d\bar{\eta}_i \,\exp\{{\bar{\eta}_k\eta^k\}}\,\Omega^\dagger\Psi .\lb{Weyl}
\eea

\item As the next step, one passes to the covariant supercharges, which act on the Hilbert space with the more natural, geometrically motivated
inner product
\bea
    \langle \Omega |\Psi\rangle =\int g\, dz\,d\bar{z}\prod_i d\eta^i\,d\bar{\eta}_i \,
    \exp\{{\bar{\eta}_k\eta^k\}}\,\Omega_{(cov)}^\dagger\,\Psi_{(cov)}.
\eea
They are related to the Weyl-ordered supercharges through the similarity transformation
\bea
    \left(\hat Q^i ,\hat{\bar{Q}}_j\right)_{cov} = g^{-{1\over 2}} \left(\hat Q^i ,\hat{\bar{Q}}_j\right) g^{{1\over 2}}\,.
\eea
\end{enumerate}

As the result of this procedure, we obtain the following quantum operators
\bea
      &&\hat Q_{(cov)}^i =\sqrt{2}\,\hat{\eta}^i \,g^{-{1\over 2}} \nabla_z \,,\qquad  \hat{\bar{Q}}_{(cov)j} =\sqrt{2}\, \hat{\bar{\eta}}_j\,
      g^{-{1\over 2}}\bar{\nabla}_{\bar{z}}  \,,\nn
      &&\hat{F} =- 2\kappa\left(\hat{z} \partial_z -\hat{\bar{z}}\partial_{\bar{z}}\right)
      -\left(2\kappa-\frac{1}{2}\right)\hat{\eta}^k \hat{\bar{\eta}}_k\,,
      \qquad
      \hat{I}^i_j =\hat{\eta}^i \hat{\bar{\eta}}_j  -\frac{1}{2}\delta_j^i\hat{\eta}^k \hat{\bar{\eta}}_k\,. \lb{gen242}
\eea
They satisfy the $su(2|1)$ superalgebra \p{salgebra} with the quantum Hamiltonian
\bea
    \hat{H} =  g^{-1}\, \bar{\nabla}_{\bar{z}}\,\nabla_z -2\kappa m\left(\hat{z} \partial_z
    -\hat{\bar{z}}\partial_{\bar{z}}\right)+m\left(1-2\kappa\right)\hat{\eta}^k\hat{\bar{\eta}}_k
    -\frac{1}{4}\,g^{-2}R\,\varepsilon_{kl}\varepsilon^{nj}\hat{\eta}^k\hat{\eta}^l\hat{\bar{\eta}}_n\hat{\bar{\eta}}_j . \lb{QuantH2}
\eea
Note that the second term in \p{QuantH2} can be re-absorbed into a redefinition of the external magnetic field in $\nabla_z, \bar\nabla_z\,$
at cost of appearance of some bosonic potential. We will explicitly do this in the next Section.

Let us also define one more $U(1)$ generator
\bea
    \hat{E}=-\left(\hat{z} \partial_z -\hat{\bar{z}}\partial_{\bar{z}}\right) -\hat{\eta}^k \hat{\bar{\eta}}_k\,.\label{ext}
\eea
It commutes with all $SU(2|1)$ generators, provided that $f(z,\bar{z})=\tilde{f}(z\bar{z})$. Thus, in this case
there is an extra $U(1)$ generator playing the role of external Casimir operator of the $su(2|1)$ superalgebra.
In the next Section we will see, on
the simple example, that the presence of this $U(1)$ generator proves crucial for finding the quantum spectrum. Note that, since both operators
$\hat{E}$ and $\hat{F}$ commute with $\hat H$, the same is true for the fermionic number operator $\hat{\eta}^k\hat{\bar{\eta}}_k$ which
is a linear combination
of these two conserved $U(1)$ generators. It is seen from \p{gen242} that at $\kappa =0$, when there are no restrictions on $f(z, \bar z)\,$,
the fermionic number operator coincides (up to the factor $1/2$) with the generator $\hat{F}\,$.
\setcounter{equation}{0}
\section{The model on a plane}
\subsection{Lagrangian and Hamiltonian}
The model on a plane corresponds to the simplest choice of the K\"ahler potential in \p{kinterm}:
\bea
    f\left(\Phi , \Phi^\dagger\right)= \Phi  \Phi^\dagger .\lb{fplane}
\eea
For this particular case, the general component Lagrangian \p{Lon} is reduced to
\bea
    {\cal L} &=& \dot{\bar{z}}\dot{z} +i m\left(2\kappa-\frac{1}{2}\right)\left(\dot{\bar{z}}z -\dot{z}\bar{z}\right)
    + \frac{i }{2}\left(\bar{\xi}_i\dot{\xi^i}-\dot{\bar{\xi}}_i\xi^i \right)+ 2\kappa \left(2\kappa -1\right) m^2 \bar{z}z\nn
    &&  + \left(1 -2\kappa \right)m\left(\bar{\xi}\cdot\xi\right).\label{PlaneL}
\eea
It is invariant under the transformations
\bea
    &&\delta z  = -\sqrt{2}\,\epsilon_i \xi^i ,\qquad\delta \xi^i =  \sqrt{2}\,i\bar{\epsilon}^i\dot z  -2\sqrt{2}\,\kappa m \bar{\epsilon}^i z\, .
\eea

In accordance with the notations of the previous Section, we will deal with the set of variables $\left(z, \eta^i\right)$
(in the considered case $\xi^i \equiv \eta^i$, since $g=1$).  The corresponding canonical Hamiltonian \p{Hcl1} is reduced to the expression
\bea
    H &=&\left[ p_z -\frac{i }{2}\left(1-4\kappa\right)m \bar{z} \right]\left[p_{\bar{z}}+\frac{i }{2}\left(1-4\kappa\right) m z \right]
     +2\kappa \left(1 -2\kappa\right) m^2 \bar{z}z\nn
     &&+\,\left(1-2\kappa\right)m\,\eta^k\bar{\eta}_k\,, \lb{AltCl}
\eea
or to the alternative expression
\bea
    H =\left( p_z -\frac{i }{2}m \bar{z} \right)\left(p_{\bar{z}}+\frac{i }{2}m z \right)
    - 2i\kappa m\left(z p_z -\bar{z}p_{\bar{z}}\right)+m\left(1-2\kappa\right)\eta^k\bar{\eta}_k\,.\label{Hclp}
\eea
Quantization is performed in the standard way
\bea
    &&[ \hat z, \hat{p}_z] =i\,, \qquad \{\hat{\eta}^i , \hat{\bar{\eta}}_j \} = \delta^i_j\,,\qquad [\hat{p}_z , \hat{\eta}^i ] =
    [\hat{p}_z , \hat{\bar{\eta}}_j ]=0\,,
    \nn &&  \hat p_z = -i\partial_z\,, \qquad \hat{\bar{\eta}}_j = \frac{\partial}{\partial \hat{\eta}^j} .
\eea
The quantum Hamiltonian
\bea
    \hat{H} = \bar{\nabla}_{\bar{z}}\,\nabla_z -2\kappa m\left(\hat{z} \partial_z
    -\hat{\bar{z}}\partial_{\bar{z}}\right)+m\left(1-2\kappa\right)\hat{\eta}^k\hat{\bar{\eta}}_k \label{HQP}
\eea
and the quantum operators
\bea
      &&\hat Q^i =\sqrt{2}\,\hat{\eta}^i \nabla_z \,,\qquad  \hat{\bar{Q}}_{j} =\sqrt{2}\, \hat{\bar{\eta}}_j\bar{\nabla}_{\bar{z}}  \,,\\
      &&\hat{F} = -2\kappa\left({z} \partial_z - {\bar{z}}\partial_{\bar{z}}\right)
      -\left(2\kappa-\frac{1}{2}\right)\hat{\eta}^k \hat{\bar{\eta}}_k\,,\qquad
      \hat{I}^i_j =\hat{\eta}^i \hat{\bar{\eta}}_j  -\frac{1}{2}\delta_j^i\hat{\eta}^k \hat{\bar{\eta}}_k\,,\lb{quantGen}
\eea
form the $su(2|1)$ superalgebra \p{salgebra}. Here,
\bea
    \nabla_z = -i\partial_z  -\frac{i }{2}m \bar{z} \,,\quad   \bar{\nabla}_{\bar{z}}=-i\partial_{\bar{z}}+\frac{i }{2}m z\,, \qquad
    \left[\nabla_z ,\bar{\nabla}_{\bar{z}}\right] = m\, .
\eea

The Hamiltonian \p{HQP} can be rewritten, up to a constant shift $2m\kappa$, in the form analogous to the classical expression \p{AltCl}
\be
 \hat{H} = -\Big[\partial_z +\frac{1}{2}(1 - 4\kappa)m\bar z\Big]\Big[\partial_{\bar z} -\frac{1}{2}(1 - 4\kappa)mz\Big] +
  2\kappa(1 -2\kappa)m^2 z\bar z + \left(1-2\kappa\right)m\,\hat{\eta}^k\hat{\bar{\eta}}_k\,. \lb{HQP1}
\ee
It is seen from this representation that we are dealing with a superextension of the two-dimensional harmonic oscillator with the
strength $ 2\kappa (1 - 2\kappa)m^2$, supplemented by a coupling to the external magnetic field
${\cal A}_z = -\frac{i}{2}(1-4\kappa)m\,\bar z\,, \;{\cal A}_{\bar z} = \frac{i}{2}(1-4\kappa)m\,z \,$.

For further use, it will be  instructive to know the explicit expressions of the $SU(2|1)$ Casimir operators defined in \p{eq413}.
For the specific realization of the quantum $SU(2|1)$ generators \p{HQP} and \p{quantGen} they are
\bea
    m^2 C_2 &= &\left(\hat H- 2\kappa m\hat{E}\right)\left(\hat H- 2\kappa m\hat{E}-m\right),\nn
    m^3C_3 & =& \left(\hat H- 2\kappa m\hat{E}\right)\left(\hat H- 2\kappa m\hat{E}-m\right)\left(\hat H- 2\kappa m\hat{E}-\frac{m}{2}\right).
    \label{Casimhat1}
\eea
Comparing these expressions with those for the $({\bf 1, 4, 3})$ oscillator model, eqs. \p{Casimhat}, we observe that they involve,
besides the Hamiltonian $\hat H$,
also the extra $U(1)$ generator $\hat{E}$ defined in \p{ext} and commuting  with all $SU(2|1)$ generators.

\subsection{Wave functions and spectrum}
It is convenient to seek the bosonic wave function $\Omega$ as an eigenfunction of the mutually commuting $U(1)$ operator \p{ext}
and the Hamiltonian \p{HQP}.
The corresponding eigenvalue problem is set by the equations
\bea
\mbox{(a)}\;\;    \hat{E}\,\Omega= n\,\Omega,\qquad
\mbox{(b)}\;\;   \hat{H}\,\Omega= {\cal E}\,\Omega = m q\,\Omega.
\eea
The equation (a) yields\footnote{We could equally choose, from the very beginning, the solution with negative $n$, $\Omega' = z^{|n|}A'(w)\,$.
The corresponding sets of wave functions are related through the complex conjugation.}
\bea
    \Omega =\bar{z}^{n}A\left(w\right)\,, \qquad w \equiv z\bar{z}\,.
\eea
Then the equation (b) amounts to the following one for $A\left(w\right)$:
\bea
    &&\left[-w \partial_w^2 -(1+n)\partial_w  +\frac{m^2}{4}w -\frac{m}{2}\right]A\left(w\right)= m\left(q - 2\kappa n + \frac{n}{2}\right)
    \,A\left(w\right).
\eea
It is solved by
\bea
    \label{Laguerre}
    A(w) =  e^{-\frac{m w}{2}}L_{q - 2\kappa  n }^{(n)}(m w),
\eea
where $L_{q - 2\kappa  n}^{(n)}(m w)$ are the generalized Laguerre polynomials. Thus the eigenvalue problem for $\hat H$ can be rewritten as
\bea
\hat{H}\,\Omega^{(\ell ; n)}= {\cal E}^{(\ell ; n)}\,\Omega^{(\ell ; n)}\,,
\eea
with
\bea
 {\cal E}^{(\ell ; n)} = m( \ell+ 2\kappa  n)
\eea
and
\bea
    \Omega^{(\ell ; n)}=\bar{z}^{n} e^{-\frac{m z\bar{z}}{2}} L_{\ell }^{(n)}(m z\bar{z})
    =\frac{z^{-n}}{\ell !} e^{\frac{m z\bar{z}}{2}} \left.\frac{d^\ell}{dw^\ell}\left(e^{-m w} w^{n+\ell}\right)\right|_{w=z\bar{z}}. \label{eigenF}
\eea
According to the definition of Laguerre polynomials, $\ell$ is a non-negative integer, $\ell \geq 0$.

The orthogonality of $\Omega^{(\ell ; n)}$ with respect to the inner product,
\bea
    \langle \Omega^{(\ell_1 ; n_1)} |\Omega^{(\ell_2 ; n_2)}\rangle := \int \, dz\,d\bar{z}\,\left(\Omega^{(\ell_1 ; n_1)}\right)^\dagger\,
    \Omega^{(\ell_2 ; n_2)}
    = \frac{\pi (n+\ell)!}{\ell ! \,m^{n+1}}\,\delta^{\ell_1 \ell_2}\,\delta^{n_1 n_2}\,, \lb{orthogon1}
\eea
is necessary for the super wave functions to form the complete orthogonal set. This orthogonality condition constrains $n$
to the integer values $n \geq -\ell\,$ \footnote{For the negative values of $n\,$, $0>n>-\ell\,$, the wave functions \p{eigenF}
remain regular at $z=0\,$.}.
The integral in \p{orthogon1} is convergent for $m>0\,$\footnote{For $m<0$, we can take advantage of the equivalent
redefinition \p{reflection} to bring all the quantum relations and formulas to the same form as for $m>0\,$, with $m \rightarrow |m|\,$.}.
The energies are positive and $\hat{H}$ \p{HQP1} is bounded from below only under the following restriction on the
parameter $\kappa$:
\be
0 \leq \kappa \leq 1/2\,.
\ee

The wave functions $\Omega^{(\ell ; n)}$ satisfy the relations
\bea
    &&\nabla_z \,\Omega^{(\ell ; n)} =i m\,\Omega^{(\ell -1; n +1)},\qquad
    \left(\nabla_z + i m \bar{z}\right)\Omega^{(\ell ;n)}=im\,\Omega^{(\ell ; n +1)},\nn
    &&\bar{\nabla}_{\bar{z}}\,\Omega^{(\ell ; n)}= -i\left(\ell +1\right)\Omega^{(\ell +1; n -1)},\qquad
    \left(\bar{\nabla}_{\bar{z}} - i m z\right)\Omega^{(\ell ; n)}=-i\left(\ell +n\right)\Omega^{(\ell ; n -1)},\nn
    &&\nabla_z \,\Omega^{(0 ; n)}=0,\qquad \left(\bar{\nabla}_{\bar{z}} - i m z\right)\Omega^{(\ell ;-\ell)}= 0\,, \lb{relOmega}
\eea
which follow from the definition \p{eigenF}. The operators $\left(\nabla_z + i m \bar{z}\right),\; \left(\bar{\nabla}_{\bar{z}} - i m z\right)$
commute with the covariant momenta $\nabla_z ,\;\bar{\nabla}_{\bar{z}} $. Using \p{relOmega}, we can obtain the convenient representation
for the generic $\Omega^{(\ell ; n)}$ as
\bea
    &&\Omega^{(\ell ; n)} =\frac{(-i)^{n}}{\ell !m^{\ell +n}}\left(\bar{\nabla}_{\bar{z}}\right)^\ell\left(\nabla_z
    + i m \bar{z}\right)^{\ell +n}\Omega^{(0; 0)}(w),
\eea
where $\Omega^{(0; 0)}(w)$ is the ground state wave function:
\bea
    \Omega^{(0 ; 0)}(w)= e^{-\frac{m z\bar{z}}{2}}\, .
\eea

Acting on $\Omega^{(\ell ; n)}$ by the supercharges $Q^i$ , we can produce all other common eigenstates of the Hamiltonian $\hat{H}$ and the external
$U(1)$ charge operator $\hat{E}$. In the process, one should take account of the  physical condition:
\bea
    \bar{\eta}_j\,\Omega^{(\ell ; n)}=0\;\Rightarrow \;\bar{Q}_j\,\Omega^{(\ell ; n)}=0\,.
\eea
Using  the relations \p{relOmega}, it is easy to find
\bea
    Q^i\,\Omega^{(\ell ; n)}=i m\sqrt{2}\, \eta^i\,\Omega^{(\ell -1; n +1)},\qquad  \varepsilon_{ij}Q^i\,Q^j\,\Omega^{(\ell ; n)}=-2 m^2 \,
    \varepsilon_{ij} \eta^i \eta^j\,\Omega^{(\ell -2; n +2)}\,.
\eea
Then the super wave functions,
\bea
    &&\Psi^{(\ell ; n)}=a^{(\ell; n)}\Omega^{(\ell;n )}+b_i^{(\ell ; n)}\eta^i\,\Omega^{(\ell -1; n +1)}
    + \frac{1}{2}\,c^{(\ell; n)}\,\varepsilon_{ij} \eta^i \eta^j \,\Omega^{(\ell -2; n +2)},\quad \ell\geq 2,\nn
    &&\Psi^{(1 ; n)}=a^{(1 ; n)}\,\Omega^{(1     ; n)}+b_i^{(1 ; n )}\eta^i\,\Omega^{(0;n +1)}, \nn
    &&\Psi^{(0 ; n)}=a^{(0; n)}\,\Omega^{(0; n)}\,,\label{WSF}
\eea
span the full Hilbert space of quantum states of the model. We observe that the ``ground states'' ($\ell =0$) and
the first excited states ($\ell =1$)
are special, in the sense that they encompass non-equal numbers of bosonic and fermionic states. The eigenvalues
of the operators $\hat{E}$ and $\hat{H}$
are given by
\bea
    \hat{E}\,\Psi^{(\ell ; n)}= n\,\Psi^{(\ell ; n)},\qquad
    \hat{H}\,\Psi^{(\ell ; n)}= {\cal E}^{(\ell ; n)}\,\Psi^{(\ell ; n)}\,, \quad
    {\cal E}^{(\ell ; n)}= m\left(2\kappa  n + \ell \right).\label{eigenV}
\eea

The ``ground states'' are annihilated by both supercharges
\be
Q^i\Omega^{(0; n)} = \bar Q_i \Omega^{(0; n)} = 0\,.
\ee
The true ground state annihilated also by $\hat{H}$ corresponds to $n=0$ or $\kappa =0, n\neq 0$. The second option shows up a
degeneracy parametrized by the number $n$ (see below). For generic $\kappa$ there is an infinite tower of the ``ground states'' parametrized by $n$,
with the energy ${\cal E}^{(0;n)} = 2\kappa m\, n\,.$ They all are annihilated by both supercharges. The combination $\hat H - m\hat F$ yields
zero on all these states, but it cannot be chosen as the ``genuine'' Hamiltonian, since it generically {\it does not} commute with the supercharges
(e.g., when acting on the states with $\ell \neq 0\,$). These surprising features of the quantum picture are in a sharp contrast with what happens
in the standard ${\cal N}=4$ SQM based on the chiral $({\bf 2, 4, 2})$ multiplet (see, e.g., \cite{IvSmi,How}).

Since for each $n$ we are dealing with finite-dimensional representations of $su(2|1)$ realized on the super wave functions,
the Casimir operators are given by the same general expression as in \p{betalambda}. Using the formulas \p{eigenV} and \p{Casimhat1},
we find that $\lambda =1/2$ for any $\Psi^{(\ell ;n)}$ and
\be
    C_2(\ell) = \left(\ell - 1 \right)\ell, \quad C_3(\ell) = \left(\ell  - 1/2 \right)\left(\ell - 1 \right)\ell ,\quad \beta(\ell) = \ell  - 1/2\,.\lb{casim4}
\ee
These values coincide with those pertinent to the oscillator model (eq. \p{Casimnumbers}). Thus in the $({\bf 2, 4, 2})$ model under consideration
the Hilbert space is spanned by the same irreps of $SU(2|1)$ as in the oscillator $({\bf 1, 4, 3})$ model. As distinct from the latter,
at any fixed level $\ell$ one finds an infinite tower of irreps parametrized by $n \geq -\ell$ and exhibiting an equidistant energy spectrum,
with spacing $2\kappa\, $.

Supercharges do not depend on $\kappa$ and, as a result, the super wave functions $\Psi^{(\ell;n)}$ involve no $\kappa$-dependent
terms in their $\eta$-expansions.
The parameter $\kappa$ is still present in the Hamiltonian \p{HQP} and in the internal $U(1)$ generator $\hat{F}$.
As was already mentioned, in the anticommutator $\{\hat{Q},\hat{\bar{Q}}\}$ there appears just the combination $\hat{H} -m\hat{F}\,$,
which involves no dependence on $\kappa$.

The norms of all super wave functions \p{WSF} are positive-definite. Using the inner product \p{orthogon1}, we define
the norms as
\be
    ||\Psi^{(\ell ; n)} ||^2 = \frac{\langle \Psi^{(\ell ; n)} |\Psi^{(\ell ; n)} \rangle}{ \langle \Omega^{(\ell ; n)} |\Omega^{(\ell ; n)} \rangle}.
\ee
This yields the following manifestly positive norms
\bea
    &&||\Psi^{(\ell ; n)} ||^2 = \bar{a}^{(\ell ; n)}a^{(\ell ; n)}+ \frac{\bar{b}^{(\ell ; n)i}b^{(\ell ; n)}_i} {m \ell}
    + \frac{\bar{c}^{(\ell ; n)}c^{(\ell ; n)}}{m^2\left(\ell -1\right)\ell},\quad \ell \geq 2 ,\nn
    &&||\Psi^{(1 ; n)} ||^2 = \bar{a}^{(1 ; n)} a^{(1 ; n)}+\frac{\bar{b}^{(1 ; n)i} b^{(1 ; n)}_i}{m},\quad  ||\Psi^{(0 ; n)} ||^2
    = \bar{a}^{(0 ; n)} a^{(0 ; n)}.
\eea

\subsection{Degeneracies}
At some special values of $\kappa$ the considered model reveals degeneracies in the energy spectrum, which amounts to the property
that the symmetry algebra  $su(2|1)$ is properly enhanced in these cases. Here we consider the cases $\kappa = 0$ and $\kappa  =1/2$,
leaving the discussion of two other, more complicated options for Appendix.

\subsubsection{$\kappa =0$}
In this case  the $su(2|1)$ superalgebra is extended by the operators $\left(\nabla_z + i m \bar{z}\right),\; \left(\bar{\nabla}_{\bar{z}} - i m z\right)$,
which commute with the Hamiltonian \p{HQP} and generate what is called the ``magnetic translation'' algebra \cite{LandPlan}. These operators also commute with all
other $su(2|1)$ generators, so we are dealing with a direct sum of $su(2|1)$ and the magnetic translation algebra. The associated degeneracy is revealed in
the property that at any level $\ell$ the energy does not depend on the parameter $n$
\bea
    {\cal E}^{(\ell ; n)} = m\,\ell .
\eea
The wave function at the energy level $\ell$ is given by the sum
\bea
    \Psi_\ell =\sum_{n = -\ell}^{\infty}s_{\ell\,n}\Psi^{(\ell ; n)}\,, \quad \hat{H}_{(\kappa = 0)}\Psi_\ell=  m\,\ell\,\Psi_\ell\,, \lb{kappa0}
\eea
where $s_{\ell\,n}$ are arbitrary coefficients. It is easy to check, that the ground state $\Psi_0$ have a simple
expression through some antiholomorphic function ${\cal S}(\bar{z})$:
\bea
    \Psi_0 =e^{-\frac{mz\bar{z}}{2}} {\cal S}(\bar{z}), \quad {\cal S}(\bar{z}) = \sum_{n = 0}^{\infty}s_{0\,n}\,\bar{z}^n \,.
\eea
{}From this representation, it immediately follows, in particular,  that
\bea
    \hat Q^i\Psi_0 = \hat{\bar{Q}}_j \Psi_0 =0\, .
\eea

\subsubsection{$\kappa =1/2$}
In this case the fermionic terms entirely drop out from the Hamiltonian \p{HQP}, so the latter becomes purely bosonic
and commuting with the fermionic
operators $\hat{\eta}^i , \hat{\bar{\eta}}_j$. It is easy to check that the $\kappa =1/2$ Hamiltonian also commutes with the operators
$\nabla_z ,\;\bar{\nabla}_{\bar{z}} \,$. The full set of the  additional bosonic and fermionic integrals of motion,
$\hat{\eta}^i , \hat{\bar{\eta}}_j, \nabla_z , \bar{\nabla}_{\bar{z}} \,$, does not commute with the rest of the $su(2|1)$ generators;
the superalgebra of these ``magnetic supertranslations'' forms a semi-direct sum with $su(2|1)$. The basic new non-vanishing (anti)commutators
of this extended inhomogeneous superalgebra are given by
\bea
&&[\hat Q^i, \bar{\nabla}_{\bar{z}}]=\sqrt{2}\,m\hat{\eta}^i\,, \quad \{\hat Q^i, \hat{\bar{\eta}}_j\} = \sqrt{2}\delta^i_j \nabla_z\,, \nn
&& [\hat{\bar{Q}}_j,\nabla_z]=-\sqrt{2}\,m\hat{\bar{\eta}}_j\,, \quad \{\hat{\bar{Q}}_i, {\eta}^j\} = \sqrt{2}\delta^j_i
\bar\nabla_{\bar z}\,. \lb{insu21}
\eea

The relevant infinite degeneracy of the energy levels is manifested  in the structure of the generic super wave function for $2\kappa =1$
\bea
    \Psi_q =\sum_{\ell = 0}^{\infty}d_{\ell \,q}\Psi^{(\ell ; q-\ell)}, \quad \hat{H}_{(\kappa =1/2)} \Psi_q = m q\, \Psi_q\,,
    \quad q=0,1,2\ldots\,,\;({\cal E}=mq)\lb{kappaone}
\eea
where $d_{\ell \,q}$ are some numerical coefficients. As distinct from the case \p{kappa0}, here the degeneracy arises between
super-wave functions belonging to different
$\ell$ levels, in accord with the property that the new symmetry generators $\hat{\eta}^i ,
\hat{\bar{\eta}}_j, \nabla_z , \bar{\nabla}_{\bar{z}} \,$
mix various terms in the sum \p{kappaone}, for instance,
$$
\hat{\eta}^i\Psi^{(\ell ; q-\ell)} \sim \Psi^{(\ell + 1; q -\ell -1)}\,, \quad \hat{\bar{\eta}}_j\Psi^{(\ell ; q-\ell)} \sim
\Psi^{(\ell - 1; q -\ell +1)}\,, \quad \hat{\bar{\eta}}_j\Psi^{(0 ; q)} = 0\,.
$$
The action of the bosonic operators $\nabla_z , \bar{\nabla}_{\bar{z}} \,$ can be found from the relations \p{relOmega}.
The $SU(2|1)$ supercharges
take each super wave function  in the sum \p{kappaone} into itself.

The ground state $\Psi_0 $ (${\cal E}=0, q=0$) in this case is
\bea
    \Psi_0 =\sum_{\ell=0}^{\infty}d_{\ell\,0}\Psi^{(\ell ; -\ell)}.\label{gr11}
\eea
Using the formula
$$
\Omega^{(\ell;-\ell)} = e^{-\frac{mz\bar z}{2}} \,\frac{(-m)^\ell}{l!}\,z^\ell\,,
$$
one can represent $\Psi^{(\ell ; -\ell)}$ for a given $\ell$ as
\bea
\Psi^{(\ell ; -\ell)}=e^{-\frac{mz\bar{z}}{2}}
    \left[ \tilde{a}_{(\ell)}\, z^{\ell} +\tilde{b}_{(\ell)\,i}\,\eta^i z^{\ell-1}+
    \tilde{c}_{(\ell)}\,\varepsilon_{ij}\, \eta^i \eta^j z^{\ell -2}\right],\lb{ll}
\eea
where $ \tilde{a}_{(\ell)},\tilde{b}_{(\ell)\,i}, \tilde{c}_{(\ell)}$ are arbitrary numerical coefficients, with the only restrictions
$\tilde{b}_{(0)\,i} = \tilde{c}_{(0)} = \tilde{c}_{(1)} =0\,$. Substituting this into the sum \p{gr11}, we can present the ground
state wave function as
\bea
 \Psi_0 = e^{-\frac{mz\bar{z}}{2}}\left[ {\cal D}_0(z)
 +{\cal D}_{1 i}(z)\,\eta^i + {\cal D}_2(z)\,\varepsilon_{ij}\, \eta^i \eta^j\right],
\eea
where ${\cal D}_0(z), {\cal D}_{1 i}(z), {\cal D}_2(z)$ are arbitrary holomorphic functions, analytic at $z=0$.

Clearly, this infinitely degenerated ground state is not annihilated by the supercharges $\hat{Q}^i, \bar{Q}_i$. Acting by the latter on the super
wave functions \p{ll}, we observe that only $\Psi^{(0;0)} = const\, e^{-\frac{mz\bar{z}}{2}}$ is vanishing under this action,
so it is the only $SU(2|1)$ singlet
ground-state wave function. For any other $\ell$ we encounter non-trivial finite-dimensional representations of $SU(2|1)$. For $\ell=1\,$,
it is an atypical
fundamental representation, with one bosonic and two fermionic vacuum states. At any $\ell \geq 2\,$, the vacuum states are grouped into
the typical multiplets, with two bosonic and two fermionic states. The Casimir operators \p{Casimhat1} take the values \p{casim4}
on all these multiplets.
Though $\hat{H}$ in \p{Casimhat1} is zero for the vacuum states, the extra $U(1)$ charge generator $\hat{E}$ is non-vanishing,
$\hat{E}\Psi^{(\ell ; -\ell)} = -\ell\,\Psi^{(\ell ; -\ell)}\,$.

\section{Conclusions and outlook}
By this paper, we initiated the systematic study of new class of the
deformed models of supersymmetric quantum mechanics, based upon the
superfield approach. We constructed $d=1$
superspace realizations of the simplest supergroup $SU(2|1)$ which
can be treated as a deformation of the ${\cal N}=4, d=1$ super
Poincar\'e symmetry by mass parameter $m$. We showed that
$SU(2|1), d=1$ supersymmetry admits off-shell realizations on the
multiplets $({\bf 1, 4, 3})$ and $({\bf 2, 4, 2})$, like its
standard ${\cal N}=4, d=1$ prototype. The relevant most general
superfield and component actions were constructed, the quantization
was performed and, in a few simple cases, the eigenvalue problems
for the relevant Hamiltonians were solved. In the $({\bf 1, 4, 3})$
case, our results basically coincide with those of ref. \cite{WS},
and we identify the weak supersymmetry models proposed there
with the  $SU(2|1)$ SQM models of single $({\bf 1, 4, 3})$
multiplet. The $SU(2|1)$ invariant off- and on-shell $({\bf 2, 4,
2})$ models are essentially new. Their basic novel features, as
compared to the standard ${\cal N}=4$ supersymmetric $({\bf 2, 4,
2})$ SQM models, are the in-built presence of WZ terms in the
component action and appearance of one more free parameter $\kappa$,
that is the $U(1)$ charge associated with the internal $U(1)$
generator ${F}$. The presence of this parameter has a salient
impact on the structure of the space of quantum states. For special
values of $\kappa$ there appear additional interesting degeneracies.
For instance, in a simple model without WZ term and $\kappa = 1/4$,
the worldline $SU(2|1)$ symmetry is enhanced to $SU(2|2)$ (see Appendix).

For the oscillator $({\bf 1, 4, 3})$ and the plane $({\bf 2, 4, 2})$
SQM models we analyzed the $SU(2|1)$ representation contents of the
space of quantum states and found that in both cases they necessarily involve at least one
atypical $SU(2|1)$ irrep, with vanishing Casimir operators and
unequal numbers of fermionic and bosonic states, apart from  the
singlet ground state (the equality is restored only in the special case of $\kappa = 1/4$).
Thus this mismatch between bosonic and
fermionic excited states, observed for the first time in \cite{WS} for the
$({\bf 1, 4, 3})$ models, seems to be a generic feature of the
$SU(2|1)$ SQM models.

Our superfield approach enables an easy construction of the
$SU(2|1)$ SQM models involving several $({\bf 1, 4, 3})$ and/or
$({\bf 2, 4, 2})$ off-shell multiplets. Besides, the rest of non-trivial
${\cal N}=4, d=1$ multiplets, namely, the multiplets $({\bf 3, 4,
1})$ and $({\bf 4, 4, 0})$, seem also to have the appropriate
$SU(2|1)$ counterparts, and it would be very interesting to consider
the corresponding SQM models. Indeed, these multiplets are naturally
described by superfields defined on the harmonic analytic ${\cal
N}=4$ superspace \cite{ILech}, which has a counterpart among
the admissible coset manifolds of the supergroup $SU(2|1)$. It is
the following coset
$$
\frac{\{Q^{i},\bar{Q}_{j},H,I^i_j,F \}}{\{Q^1, \bar{Q}_2, F, I^1_2, I^1_1 \}}\, \sim \{Q^2, \bar{Q}_1, H, I^2_1 \}\,.
$$
The corresponding coset coordinates include half the original $\theta$
coordinates, the time $t$, and additional harmonic coordinates of
the complex internal coset $SU(2)/\{I^1_2, I^1_1\}\,$.

Besides setting up new $SU(2|1)$ SQM models along these lines and analyzing their
hidden links with the ``$N$-fold'' supersymmetries \cite{Nfold}, quasi-exactly solvable models \cite{quasi}
(a possible relation between the latter and weak supersymmetry was noticed in \cite{WS}), as well as
the higher-dimensional models exhibiting curved rigid supersymmetries, there is
one more intriguing problem for the future study. It would be tempting to extend our $d=1$ superspace
formalism to some higher-rank supergroups as the curved
analogs of higher ${\cal N}$ one-dimensional Poincar\'e supersymmetries \p{Nsusy}.
The natural choice is the supergroup $SU(2|2)$ extending $SU(2|1)$. It involves eight
supercharges and so is the appropriate candidate for deformed ${\cal N}=8, d=1$
supersymmetry. The closure of the $SU(2|2)$ supercharges contains two
commuting $SU(2)$ subalgebras and so this supergroup admits as its super cosets, besides the standard
harmonic analytic superspace, also analogs of the bi-harmonic analytic
superspaces \cite{biharm}.

Actually, this supergroup allows three
independent central charges, and two of them can be identified with
two light-cone projections of the $d=2$ translation operator. Thus
such a centrally-extended $SU(2|2)$ could also be employed as a kind of $d=2$ ``weak
supersymmetry'', and the question is whether one can construct
non-trivial $d=2$ sigma models based on such a deformation of the
flat ${\cal N}=(4, 4)$ $d=2$ supersymmetry. The problem
of generalizing the $d=1$ weak supersymmetry to $d=2$ was posed in
\cite{WS} \footnote{The possible existence of such unusual $d=2$ supersymmetric systems does not contradict
the renowned Coleman-Mandula and Haag-Lopushanski-Sohnius no-go
theorems, as the latter do not apply to one and two dimensions.}. We would like to point out that various versions of the
$SU(2|2)$ supersymmetry already appeared in the literature as the worldvolume on-shell symmetry of the Pohlmeyer-reduced $AdS_3\times S^3$
and $AdS_5\times S^5$ superstrings \cite{PR}, as well as the worldline on-shell symmetry of ${\cal N}=4$ supersymmetric
Landau problem \cite{Bych}. The relevant off-shell superfield formalism could help in getting further insights into the symmetry structure
of these and similar $d=1,2$ theories of current interest.

\section*{Acknowledgements}
We acknowledge support from the RFBR grants Nr.12-02-00517 and
Nr.11-02-90445. We are indebted to Sergey Fedoruk and Mikhail Goykhman for interest in the work. E.I. thanks Andrei Smilga
for a discussion of weak supersymmetry.
\appendix

\section{More on degeneracies in the model on a plane}

\subsection{$\kappa =1/4$}
This case is distinguished in that the WZ term in the Lagrangian \p{PlaneL} and, respectively, the coupling to the magnetic
field in the Hamiltonian \p{HQP1},
disappear:
\bea
&&{\cal L}_{(\kappa =1/4)} = \dot{\bar{z}}\dot{z}
    + \frac{i }{2}\left(\bar{\xi}_i\dot{\xi^i}-\dot{\bar{\xi}}_i\xi^i \right) -\frac14 m^2 \bar{z}z
    + \frac12 m\left(\bar{\xi}\cdot\xi\right),\lb{Lagr14} \\
&&\hat{H}_{(\kappa =1/4)} = -\partial_{\bar{z}}\,\partial_z +\frac{m^2}{4}\,\bar{z}z+\frac{m}{2}\,\hat{\eta}^k\hat{\bar{\eta}}_k \,.\lb{H14}
\eea

The Hamiltonian \p{H14} is a fermionic extension of the two-dimensional oscillator Hamiltonian. It commutes with the operators
\bea
    \hat{F}_+=\frac{1}{m}\nabla_z\left(\nabla_z + i m \bar{z}\right),\qquad \hat{F}_-=
    \frac{1}{m}\bar{\nabla}_{\bar{z}}\left(\bar{\nabla}_{\bar{z}} - i m z \right).\label{osc1}
\eea
Together with $\hat{F} = -(z\partial_z - \bar z\partial_{\bar z})$, they form the new algebra $su(2)$,
which commutes with the original $su(2)$ generators $\hat{I}_j^i$. This extra
$su(2)$ algebra is none other than the well known hidden $su(2)$ symmetry algebra
of the two-dimensional harmonic oscillator \cite{Perel}.

Acting by the new $su(2)$ generators on the $SU(2|1)$ supercharges, we obtain the new complex doublet of supercharges
$\hat S^i, \hat{\bar{S}}_j$:
\bea
    \hat S^i =\sqrt{2}\,\hat{\eta}^i \left(\bar{\nabla}_{\bar{z}} - i m z \right) \,,\qquad  \hat{\bar{S}}_{j} =\sqrt{2}\,
    \hat{\bar{\eta}}_j\left(\nabla_z + i m \bar{z}\right).\label{osc2}
\eea
The operators \p{osc1}, \p{osc2} extend the superalgebra $su(2|1)$ to the centrally extended superalgebra $su(2|2)$, with $\hat H$
as the central charge generator:
\bea
    &&\lbrace\hat{Q}^i,\hat{\bar{Q}}_j\rbrace =2\delta_j^i \hat H + 2m\left(  \hat{I}_j^i - \delta_j^i \hat{F}\right),
    \qquad\lbrace\hat{Q}^i,\hat{\bar{S}}_j\rbrace =2 m\delta_j^i \hat{F}_+,\nn
    &&\lbrace\hat{S}^i,\hat{\bar{S}}_j\rbrace =2\delta_j^i \hat H + 2m\left(  \hat{I}_j^i + \delta_j^i \hat{F}\right),
    \qquad\lbrace\hat{S}^i,\hat{\bar{Q}}_j\rbrace =2 m\delta_j^i \hat{F}_-,\nn
    &&\left[\hat{I}^i_j, \hat{\bar{Q}}_{l}\right] =\frac{1}{2}\delta^i_j\hat{\bar{Q}}_{l} -\delta^i_l\hat{\bar{Q}}_{j}\, ,\qquad
    \left[\hat{I}^i_j,\hat{Q}^{k}\right] = \delta^k_j \hat{Q}^{i} - \frac{1}{2}\delta^i_j \hat{Q}^{k},\nn
    &&\left[\hat{I}^i_j, \hat{\bar{S}}_{l}\right] =\frac{1}{2}\delta^i_j\hat{\bar{S}}_{l} -\delta^i_l\hat{\bar{S}}_{j}\, ,\qquad
    \left[\hat{I}^i_j,\hat{S}^{k}\right] = \delta^k_j \hat{S}^{i} - \frac{1}{2}\delta^i_j \hat{S}^{k},\nn
    &&\left[\hat{F}, \hat{\bar{Q}}_{l}\right] =-\frac{1}{2}\hat{\bar{Q}}_{l}\,,\qquad \left[\hat{F}, \hat{Q}^{k}\right]=\frac{1}{2}\hat{Q}^{k},
    \qquad \left[\hat{F}_+, \hat{\bar{Q}}_{l}\right] =\hat{\bar{S}}_l\,,\qquad\left[\hat{F}_-, \hat{S}^i\right] =-\hat{S}^i,\nn
    &&\left[\hat{F}, \hat{\bar{S}}_{l}\right] =\frac{1}{2}\hat{\bar{S}}_{l}\,,\qquad \left[\hat{F}, \hat{S}^{k}\right]=-\frac{1}{2}\hat{S}^{k},
    \qquad \left[\hat{F}_-, \hat{\bar{S}}_{l}\right] =\hat{\bar{Q}}_l\,,\qquad\left[\hat{F}_-, \hat{Q}^i\right] =-\hat{Q}^i,\nn
    &&\left[\hat{I}^i_j,  \hat{I}^k_l\right] =  \delta^k_j \hat{I}^i_l- \delta^i_l \hat{I}^k_j \,,\qquad \left[\hat{F}_+,\hat{F}_-\right]
    =2\hat{F},\qquad \left[\hat{F},\hat{F}_\pm\right]=\pm\hat{F}_\pm\,.
\eea
The extra $U(1)$ generator $\hat{E}$ defines an outer automorphism of the extended algebra,
\bea
    \left[\hat{E}, \hat{\bar{S}}_{l}\right] =2\hat{\bar{S}}_{l}\,,\qquad \left[\hat{E}, \hat{S}^{k}\right]=
    -2\hat{S}^{k},\qquad \left[\hat{E},\hat{F}_\pm\right]=\pm 2\hat{F}_\pm\,.
\eea

At any fixed energy ${\cal E}_q = mq = m(\ell +n/2)$, the numbers $n$ and $\ell$ take the values indicated in the Table
\begin{center}
$
\begin{array}{c|c|c|c|c|c|c|c}
    n & 2q & 2q-2 &\ldots & 0 & \ldots & -2q+2 & -2q\\
\hline
    \ell & 0 & 1  &\ldots & q & \ldots & 2q -1 & 2q
\end{array}\,,. $
\end{center}
The number $q = \ell +n/2$ can be non-negative integer or half-integer:
\be
     q= 0, 1/2, 1, 3/2\ldots\, .
\ee
The super wave function $\Psi_q$ with the energy ${\cal E}_q$ is given by the finite sum
\bea
    \Psi_q =\sum_{\ell=0}^{2q}r_{\ell\,q} \Psi^{(\ell; 2q-2\ell)} ,
\eea
with $r_{\ell \,q}$ being some coefficients.
This means that the super wave function $\Psi_q$ has a finite degeneracy.
The ground state (${\cal E} = 0, q=0$) is a $su(2|2)$ singlet. The excited states $\Psi_q$, with
\be
    q = 1/2, 1, 3/2\ldots ,
\ee
are combined into the $SU(2|2)$ multiplets of dimension $8q$, whence the degeneracy $8q$.

This multiplet structure and degeneracy become manifest, when counting the numbers of states in the $\eta$-expansion of $\Psi_q$:
\begin{center}
$\begin{array}{c|c|c|c}
    \eta-{\textit{monomial}}&1 & \eta & \eta^2 \\
\hline
    {\textit{degeneracy}}&2q+1 & 4q & 2q-1
\end{array}\,. $
\end{center}

Each excited energy level has an equal number of bosonic and fermionic states which form some ``short'' $SU(2|2)$
multiplet. Recall that such multiplets are characterized by the ``triple'' \cite{Beis}
\bea
\langle n_1, n_2,\vec{C}\rangle \,,
\eea
where $n_1, n_2$ are some positive integer numbers and $\vec{C}$ is a three vector with the three admissible $SU(2|2)$ central charges
as the components. The dimensionality of such a short multiplet is given by the formula
\bea
d = 4(n_1 + 1)(n_2 + 1) + 4n_1n_2\,. \lb{dimens}
\eea
Our case with one central charge corresponds to $n_1 = 2q-1, n_2 =0$ and $\vec{C}=\left(\hat{H}/m,0,0\right)$, i.e. to the triple
\bea
    \langle 2 q -1, 0,\vec{C}\rangle ,\qquad\vec{C}=(q,0,0).
\eea
Eq. \p{dimens} implies just $d=8q$ for these short  $SU(2|2)$ multiplets.

The simplest such multiplet, with $q =1/2$, encompasses two super wave functions,
$$
(\Psi^{(0;1)}\,, \;\Psi^{(1;-1)})\,,
$$
and so involves two bosonic and two fermionic states. It is a sum of the singlet and an atypical  fundamental
$su(2|1)$ multiplets. The additional supercharges $S^i$ and $\bar{S}_j$ transform these wave functions into each other:
$$
S^i\Psi^{(0;1)} \sim \Psi^{(1;-1)}\,, \; S^i\Psi^{(1;-1)} = 0\,, \quad  \bar S^i\Psi^{(0;1)} = 0\,, \;
\bar S^i\Psi^{(1;-1)} \sim \Psi^{(0;1)}\,.
$$

Our last observation concerning the $\kappa =1/4$ case is as follows. Define the new supercharges
\bea
    \hat{\Pi}^i =\hat{Q}^i+ \varepsilon^{ij}\hat{\bar{S}}_j,\qquad \hat{\bar{\Pi}}_i =\hat{\bar{Q}}_i- \varepsilon_{ij}\hat{S}^j.
\eea
One can check that they form ${\cal N}=4$ Poincar\'e superalgebra,
\bea
    \left\{\hat{\Pi}^i,\hat{\bar{\Pi}}_j \right\}=4\delta^i_j \hat{H}\,, \quad \left\{\hat{\Pi}^i,\hat{\Pi}^j \right\} =0\,,
\eea
which is a subalgebra of the centrally-extended $su(2|2)$, along with $su(2|1)\,$.  This means that the Hamiltonian \p{H14} possesses the
standard ${\cal N}=4, d=1$ supersymmetry too.

The same phenomenon manifests itself as the property that the on-shell Lagrangian \p{Lagr14}
can be equivalently constructed by eliminating the auxiliary fields in the simple ${\cal N}=4, d=1$ superfield action
of the standard chiral ${\cal N}=4, d=1$ multiplet $({\bf 2, 4, 2})$:
\bea
    {\cal L}=\frac{1}{4}\int d^2\tilde{\theta}\, d^2\tilde{\bar{\theta}}\, \tilde{\Phi}\, \tilde{\Phi}^\dagger
    +\frac{m}{2}\left[\int d^2\tilde{\theta} \,\tilde{\Phi}^{2}+ \mbox{c.c.}\right],
\eea
where $\tilde{\Phi}, \tilde{\Phi}^\dagger$ are left and right chiral superfields.
Commuting $su(2|1)$ superalgebra and ${\cal N}=4$ Poincar\'e superalgebra realized on the same on-shell set $(z, \bar z, \eta^i, \bar\eta_i)$,
we recover the centrally-extended $su(2|2)$ superalgebra as the closure of these two subalgebras.

Reversing the argument, one can say that the simplest ${\cal N}=4$ superextension of the two-dimensional harmonic oscillator
possesses hidden symmetries $su(2|1)$
and $su(2|2)$. We have failed to find such a statement in the literature. For the time being, we do not know whether
this surprising duality extends off shell.

\subsection{General rational $\kappa$}
Finally, we briefly consider the case when  $\kappa$ takes the rational values within the range $0 < \kappa < 1/2$.
Such rational $\kappa$ can be represented as $\kappa = a/2b$, where $a$ and $b>a$ are  positive integers.
The energies take the values:
\bea
    \hat{H}_{(\kappa = a/2b)}\Psi_q=  {\cal E}\,\Psi_{\cal E}=m q\,\Psi_q\,,\quad q = \frac{1}{b}(an +  b\ell)\geq 0\,.
\eea
Shifting $n$ and $\ell$ as
\bea
    n\rightarrow n\pm b\,,\qquad \ell\rightarrow \ell \mp a\,,\label{shift}
\eea
one keeps the energy intact. Making such a shift twice or more times, we always obtain the same energy $mq$.
According to the relations \p{relOmega}, such shifts can be accomplished by the successive action on the super wave function $\Psi^{(\ell;n)}$ by
the operators which are composed as the proper products of the operators
\bea
\nabla_z\,, \quad \left(\nabla_z + i m \bar{z}\right), \quad \bar{\nabla}_{\bar{z}}\,, \quad
    \left(\bar{\nabla}_{\bar{z}} - i m z \right),
\eea
and commute with $\hat{H}_{(\kappa = a/2b)}$. These products are uniquely found to be the powers of the following ``elementary'' operators
\bea
    &&J_{(ab)} = \left(\nabla_z\right)^{a}\left(\nabla_z + i m \bar{z}\right)^{b-a},\qquad \bar{J}_{(ab)} = \left(\bar{\nabla}_{\bar{z}}\right)^{a}
    \left(\bar{\nabla}_{\bar{z}} - i m z \right)^{b-a}, \quad  b> a> 0\,,  \nn
&& J_{(ab)}\Psi^{(\ell;n)} \sim \Psi^{(\ell-a; n + b)} \,, \quad  \bar{J}_{(ab)}\Psi^{(\ell;n)} \sim \Psi^{(\ell + a;n - b)}\,. \lb{actJ}
\eea

Note that the successive action of the operators $J, \bar J$ on $\Psi^{(\ell;n)}$ according to \p{actJ} cannot  take the numbers $\ell$ and $n$
out of the range of their definition, i.e. $\ell \geq 0, n \geq -\ell\,$. This can be checked, using the relations \p{relOmega}.
For the boundary values $a =0, \kappa = 0$ and $a = b, \kappa = 1/2$, these operators become the powers of the more elementary operators
$\left(\nabla_z + i m \bar{z}\right), \;\left(\bar{\nabla}_{\bar{z}} - i m z \right)$ and $\nabla_z, \bar{\nabla}_{\bar{z}}\,$, which belong
to the extended symmetry algebras of these special cases. For $\kappa = 1/4$ such elementary generators are just $\hat{F}_\pm$, eq. \p{osc1},
and they correspond to the simplest suitable choice $a=1, b =2\,$. In the generic case the operators $J_{ab}, \bar J_{ab}$ cannot be reduced
to the more elementary operators commuting with $\hat{H}$, and so they constitute nonlinear symmetry algebras. Commuting
them with the $su(2|1)$ generators,
we obtain some nonlinear extensions of $su(2|1)\,$\footnote{A nonlinear extension of $su(2|2)$ in the quantum-mechanical context was,
e.g., considered
in \cite{Plyu}.}.

\section{${\cal N}=2$ superfield formulation}
Sometimes, it is advantageous to reformulate a model with ${\cal N}$ extended supersymmetry in terms of superfields
of the lower ${\cal N}$ supersymmetry.
The models with deformed ${\cal N}=4$ supermultiplets $({\bf 1, 4, 3})$, $({\bf 2, 4,2})$ can also be described
in terms of ${\cal N}=2$, $d=1$ superfields, so
that only some ${\cal N}=2$ supersymmetry subgroup of the full $SU(2|1)$ supergroup remains manifest. In particular,
$({\bf 1, 4, 3})$ amounts to the two
${\cal N}=2$ multiplets with the off-shell contents $({\bf 1, 2, 1})$ and
$({\bf 0, 2, 2})$, while the chiral multiplet $({\bf 2, 4, 2})$ is represented by the ${\cal N}=2$ chiral multiplets $({\bf 2, 2, 0})$ and
$({\bf 0, 2, 2})$. These decompositions are similar to those for the standard ${\cal N}=4, d=1$ supermultiplets $({\bf 1, 4, 3})$
and $({\bf 2, 4, 2})$
(see, e.g., \cite{IvSmi}), but the transformation laws of the relevant ${\cal N}=2$ multiplets, even with respect
to the manifest ${\cal N}=2$ supersymmetry,
prove to be deformed by terms $\sim m\,$.

It will be convenient to choose as the ${\cal N}=2$ supercharges $Q^1 \equiv Q$ and $\bar{Q}_1 \equiv \bar{Q}$, so that
\bea
    \lbrace Q, \bar{Q}\rbrace = 2H_1\,,\qquad \left[ H_1 , Q\right]=\left[H_1 , \bar{Q}\right] = 0 ,\qquad H_1 := H + m\left( I^1_1 - F\right).
    \label{algebraN2}
\eea
The relevant Hamiltonian $H_1$ is shifted, relative to the canonical Hamiltonian $H$, by the $U(1)$ charge $\sim I^1_1 - F$. The latter commutes
with  both supercharges and so is a central charge from the ${\cal N}=2$ subalgebra point of view; it becomes, however,  ``active'' when applied
to other objects, e.g., the generators of the second ${\cal N}=2$ subalgebra $(Q^2, \bar{Q}_2)$ \footnote{This activation
of the central charge is reminiscent of the similar effect observed in some supersymmetric $d=4$ sigma models \cite{JimG}.}. The orthogonal
combination $\sim I^1_1 + F$ defines the standard $U(1)$ automorphism of the ${\cal N}=2$ superalgebra \p{algebraN2}, completing it to $u(1|1)$.

\subsection{The multiplet $({\bf 1, 4, 3})$}
The passing to the ${\cal N}=2$ superfield notation is rendered by the following redefinitions
\bea
    \theta_i \equiv \left(\theta , \vartheta\right), \quad \epsilon_i \equiv \left(\epsilon , \varepsilon\right),\quad
    Q^i \equiv \left(Q , {\cal Q}\right),\quad \psi^i \equiv \left(\chi, \psi\right),\qquad \mbox{and\; c.c.}.
\eea
The manifest ${\cal N}=2$, $d=1$ supersymmetry acts as follows
\bea
    &&\delta\theta =\epsilon, \qquad \delta\bar{\theta}=\bar{\epsilon},\qquad
    \delta t=i\left(\epsilon\,\bar\theta + \bar{\epsilon}\,\theta\right),\nn
    &&\delta\vartheta =2m\,(\bar{\epsilon}\,\theta)\,\vartheta , \qquad \delta\bar{\vartheta}=-2m\,(\epsilon\,\bar{\theta})\,\bar{\vartheta},
\eea
while the rest of the $SU(2|1)$ transformations \p{tr} as
\bea
 &&\delta\vartheta =\varepsilon, \qquad \delta\bar{\vartheta}=\bar{\varepsilon},\qquad
    \delta t=i\left(\varepsilon\,\bar\vartheta + \bar{\varepsilon}\,\vartheta\right),\nn
    &&\delta\theta =2m\,(\bar{\varepsilon}\,\vartheta)\,\theta , \qquad \delta\bar{\theta}=-2m\,(\varepsilon\,\bar{\vartheta})\,\bar{\theta}\,.
\eea
We see that the set $(t, \theta, \bar\theta)$ is closed under the standard realization of ${\cal N}=2, d=1$ supersymmetry, which, however,
non-trivially acts on the extra Grassmann coordinates $\vartheta, \bar\vartheta$. The closure of two ${\cal N}=2$ transformations on
these coordinates yields the phase $U(1)$ transformation generated just by the ``central charge'' generator appearing in \p{algebraN2}.

Now, using the standard expressions for the ${\cal N}=2$ covariant derivatives ${\cal D}, \bar{\cal D}$,
\bea
    {\cal D}=\frac{\partial}{\partial\theta}-i\bar{\theta}\,\frac{\partial}{\partial t}\,,\qquad
    \bar{{\cal D}} = -\frac{\partial}{\partial \bar{\theta}}
    +i\theta\,\frac{\partial}{\partial t}\,.
\eea
it is rather straightforward to show that the real $SU(2|1)$ superfield $G$ defined in \eqref{Gstruct} is rewritten
in the ${\cal N}=2$ superfield notation as
\bea
    G=X + \left[1-2m\bar{\theta}\theta\right]\left(\vartheta\,\bar{\Psi} -\bar{\vartheta}\,\Psi\right)
    + \vartheta\bar{\vartheta}\left[1-2m\bar{\theta}\theta\right]\left(\frac{1}{2}\left[\bar{{\cal D}},{\cal D}\right] X + 2m X\right).
\eea
Here the unconstrained bosonic ${\cal N}=2$ superfield $X(t, \theta, \bar\theta)$ comprises the ${\cal N}=2$ multiplet $({\bf 1, 2, 1})$
\bea
    X=x + \theta\,\chi -\bar{\theta}\,\bar{\chi}
    + \theta\bar{\theta}\left(A+mx\right),\qquad \delta_{\epsilon} X \simeq X'(t',\theta', \bar\theta') - X(t,\theta, \bar\theta)= 0\,, \lb{XN2}
\eea
while $\Psi, \bar\Psi$ are (anti)chiral fermionic  ${\cal N}=2$ superfields describing the multiplet $({\bf 0, 2, 2})$:
\bea
&& \bar{\cal D}{\Psi} =0\,, \qquad \Psi = \bar\psi + \theta B -i\theta\bar\theta \dot{\bar{\psi}}\,, \lb{Psiexp}\\
&& \delta_\epsilon \Psi = 2m\,(\bar{\epsilon}\,\theta) \Psi\,,  \qquad \delta_\epsilon \bar{\Psi} =
- 2m \,(\epsilon\,\bar{\theta})\,\bar{\Psi}\,.\label{trPsi}
\eea
The real and complex auxiliary fields $A$ and $B$ in the $\theta$ expansions in \p{XN2} and \p{Psiexp} are related to the
components of the original triplet auxiliary field $B^i_j$. We see that the chiral fermionic $({\bf 0, 2, 2})$ superfields $\Psi$ and
$\bar\Psi$ are transformed in the ${\cal N}=2$ supersymmetry with non-trivial weight factors which disappear only in the limit $m=0\,$.
The closure of two ${\cal N}=2$ transformations on $\Psi$ and $\bar\Psi$ yield, besides the standard $t$ shift generated by the canonical part $H$
of $H_1$ in \p{algebraN2}, also a non-trivial phase transformation generated by the central charge part which is activated on these superfields.

The second pair of supercharges ${\cal Q}$, $\bar{{\cal Q}}\,$, which correspond to the hidden supersymmetry, generates
the transformations:
\bea
    \delta_{\varepsilon} X = (1-2m\bar{\theta}\theta)(\bar{\varepsilon}\,\Psi -\varepsilon\,\bar{\Psi}) ,\qquad
    \delta_{\varepsilon} \bar{\Psi} =\bar{\varepsilon}\,{\cal D}\,(\bar{{\cal D}} -2 m\theta) X.\label{N2Hid}
\eea

The simplest oscillator model action corresponding to the choice $f(G) = \frac14 G^2$ in \p{1}, is written through ${\cal N}=2$ superfields as
\bea
    S=-\frac{1}{4}\int d\mu\, G^2  =\frac{1}{2}\int dt\, d\bar{\theta}\, d\theta\left[-\frac{1}{2}\,X\left[\bar{{\cal D}},{\cal D}\right] X - m X^2
     + \Psi\bar{\Psi} + 2m \theta\bar{\theta}\, \Psi\bar{\Psi}\right].\lb{N2act}
\eea
It is invariant under the superfield transformations \eqref{N2Hid} and \p{XN2}, \p{trPsi}. The invariance under \p{trPsi} is ensured just
due to the last term $2m \theta\bar{\theta}\, \Psi\bar{\Psi}$ which deforms the conventional ${\cal N}=2$ superfield action of
the ${\cal N}=2$ multiplets $({\bf 1,2,1})$ and $({\bf 0, 2, 2})$. This term is also responsible for the shift of the canonical Hamiltonian $\hat{H}$
with respect to the Hamiltonian $\hat{H}_1$ defined as the square of the ${\cal N}=2$ charges obeying \p{algebraN2}:
\bea
\hat H = \hat H_1 + m\,\hat{\psi}\hat{\bar{\psi}}\,, \quad
\hat{H}_1 = \frac{1}{2}\, \left( \hat p + {i m \hat x} \right)\left( \hat p - {i m \hat x} \right) +m\,\hat{\chi}\hat{\bar{\chi}}. \label{qHam1}
\eea
Thus we conclude that in the considered model not only the hidden supersymmetry, but also the manifest ${\cal N}=2$ supersymmetry undergo
a deformation disappearing only in the limit $m=0$. This is generically true for the $SU(2|1)$ models of self-interacting
$({\bf 1,4,3})$ multiplet, as well
as for the $({\bf 2,4,2})$ models.
\subsection{Degeneracy of quantum states: ${\cal N}=2$ supersymmetry view}
It is instructive to compare the spectrum of the Hamiltonians $\hat{H}$ and $\hat{H}_1$ on the same set of quantum states of the $SU(2|1)$
oscillator model with the action \p{N2act}.
Both spectra are characterized by the same linear dependence on the Landau level number $\ell$,
$$ E(\ell)=E_1(\ell) = m\ell\,.$$
However, they reveal two different patterns of degeneracy as is shown in the figure 1.
\begin{figure}[H]
\hspace*{0.1\textwidth}\includegraphics[width=0.35\textwidth]{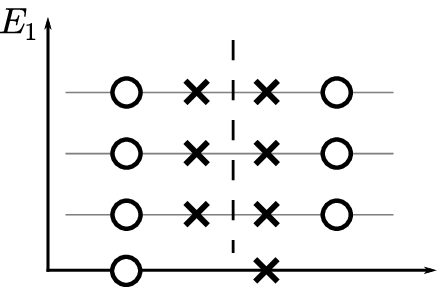}\hfill
\includegraphics[width=0.35\textwidth]{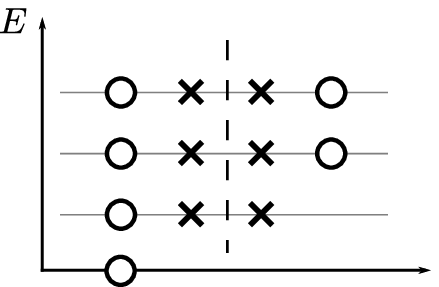}\hspace*{0.1\textwidth}\\
\hspace*{0.1\textwidth}\parbox{0.35\textwidth}{\centering (a) The spectrum of $\hat{H}_1$.}
\hfill
\parbox{0.35\textwidth}{\centering (b) The spectrum of $\hat{H}$.}\hspace*{0.1\textwidth}\\
\caption{The degeneracy of Landau levels. Circles and crosses indicate bosonic and fermionic states.}
\label{fig}
\end{figure}
We see that the spectrum of $\hat{H}_1$ is split into two towers of ${\cal N}=2$ supersymmetry multiplets, such that
the left and right towers are built on the bosonic and fermionic zero-energy ground states, respectively. These ground states are  ${\cal N}=2$
singlets, while the excited states reveal double degeneracy, as in any standard ${\cal N}=2$ SQM. The additional degeneracy between the states
from the left and right  towers is related to the fact that  $\hat{H}_1$, as well as the ${\cal N}=2$ supercharges $\hat{Q} = \hat\chi(\hat p -imx),
\hat{\bar{Q}} = \hat{\bar\chi} (\hat p + i mx)$, commute with the additional fermionic translation generators $\hat{\psi}$, $\hat{\bar{\psi}}$
(which do not belong to the $su(2|1)$ superalgebra)\footnote{The origin of this degeneracy can be explained as follows. The Hamiltonian $H_1$
can be alternatively viewed as the canonical one for the action \p{N2act}, in which the last term $\sim \theta\bar\theta$ is suppressed.
Such truncated ${\cal N}=2$ action reveals an invariance under the purely fermionic shifts $\Psi \rightarrow \Psi + \omega\,,$ which
is responsible for the degeneracy just mentioned. Neither the full action \p{N2act} nor the genuine Hamiltonian $\hat{H}$ possess
such an extra ``accidental'' invariance.}. They also commute with the $U(1)$ generator $\hat{\psi}\hat{\bar{\psi}} = \hat{F} - \hat{I}^1_1$ which
takes the eigenvalues $0$ and $1$ on all states from the left and the right towers, respectively. The ${\cal N}=2$ superalgebra automorphism
generator $\hat{F} + \hat{I}^1_1 = \hat\chi\hat{\bar{\chi}}$ is zero on both ground states and discriminates the fermionic and bosonic states
at the excited levels inside each tower.

As regards the spectrum of $\hat{H}$, the relation \p{qHam1} between
$\hat{H}$ and $\hat{H}_1$ and the fact that all states in the right
tower are eigenfunction of $\hat{\psi}\hat{\bar{\psi}}$ with the
eigenvalue $+1$ imply that this tower is lifted up just by one level
compared to Fig 1(a). At the same time, the left tower remains
unaffected because $\hat{\psi}\hat{\bar{\psi}}$ is zero on all its
states. Thus, the hidden supersymmetry is restored and the former
fermionic vacuum state combines with the first-level excited ${\cal
N}=2$ states, forming together three excited $\ell=1$ states
belonging to the fundamental representation of $SU(2|1)\,$. The
action of the $U(1)$ generators $\hat{F} \pm \hat{I}^1_1\,$, as well
as of the ${\cal N}=2$ supercharges $\hat{Q}, \hat{\bar{Q}}\,$, on
all states does not change. The degeneracy between the excited levels from
the left and right towers is now due to the hidden supersymmetry, generators of which commute with $\hat H$
(but not with $\hat H_1$).

\subsection{On the multiplet $({\bf 2, 4, 2})$}
This $SU(2|1)$ multiplet is described by the chiral bosonic and fermionic ${\cal N}=2$ superfields $Z$ and $\Pi$ representing, respectively, the
multiplets $({\bf 2, 2, 0})$ and $({\bf 0, 2, 2})$
\bea
    && Z=z + \sqrt{2}\,\theta\,\xi - i\theta\bar{\theta}\,\dot{z},\qquad \bar{{\cal D}}Z=0,\nn
    && \Pi =\pi + \sqrt{2}\,\theta B- i\theta\bar{\theta}\,\dot{\pi},\qquad \bar{{\cal D}}\Pi =0 ,
\eea
where $\xi^i\equiv\left(\xi , \pi\right)$, $\bar{\xi}_i\equiv\left(\bar{\xi} , \bar{\pi}\right)$.
Like in the previous case,   under ${\cal N}=2$ supersymmetry these superfields transform with non-trivial weights $\propto m$:
\bea
    \delta_{\epsilon}\Pi =2m\left(2\kappa - 1\right)(\bar{\epsilon}\,\theta)\,\Pi,\qquad
    \delta_{\epsilon}Z=4m\kappa\,(\bar{\epsilon}\,\theta)\,Z. \lb{N2ZPi}
\eea
The hidden supersymmetry acts on them as follows:
\bea
    \delta_{\varepsilon} Z = -\sqrt{2}\,\varepsilon\,\Pi ,\qquad
    \delta_{\varepsilon} \Pi =\frac{\sqrt{2}}{2}\,\bar{\varepsilon}\,\bar{{\cal D}} \left[(1 + 2m\theta\bar\theta)DZ\right]
    -2\sqrt{2}\,m \kappa\,\bar{\varepsilon}\,Z . \lb{N2ZPi2}
\eea

The simplest $({\bf 2, 4, 2})$ Lagrangian corresponding to the choice \p{fplane} in \p{kinterm} is rewritten
in the ${\cal N}=2$ superfield notation as
\bea
    {\cal L}&=&\int d\bar{\theta}\, d\theta\,
    \bigg\lbrace \frac{1}{4}\left(1 + 4\kappa m {\theta}\bar\theta\right)\bar D\bar Z DZ  + \frac{m}{2}(4\kappa -1)\,Z\bar Z +
     2m^2\kappa(2\kappa -1)\theta\bar\theta\,Z\bar Z\nn
    && +\,\frac{1}{2}\left[1 + 2m\left(2\kappa - 1\right) \theta\bar{\theta}\right] \Pi\bar{\Pi} \bigg\rbrace .\label{N2ch}
\eea
It is invariant, up to a total derivative, under \p{N2ZPi} and \p{N2ZPi2}. Thus here we again encounter the deformed transformations and
Lagrangian even at the level of ${\cal N}=2$ supersymmetry. The $\theta$ dependent terms cannot be removed from \p{N2ZPi} and
\p{N2ch} by any choice of the parameter $\kappa$. They disappear  in the $m=0$ limit only. Note that the Lagrangian \p{N2ch} simplifies
for the special values of $\kappa =0, 1/2, 1/4$:
\bea
&&{\cal L}_{(\kappa =0)} = \int d\bar{\theta}\, d\theta\, \Big[ \frac{1}{4}\,\bar D\bar Z DZ -\frac{m}{2}Z\bar Z
+ \frac12(1 - 2m\theta\bar\theta)\Pi\bar\Pi \Big],  \nn
&&{\cal L}_{(\kappa =1/2)} = \int d\bar{\theta}\, d\theta\, \Big[ \frac{i}{4}(1 +2m\theta\bar\theta)
\left(Z\dot{{\bar Z}} - \dot{Z}\bar Z\right)+ \frac12\Pi\bar\Pi \Big],  \nn
&&{\cal L}_{(\kappa =1/4)} = \int d\bar{\theta}\, d\theta\, \Big[ \frac{1}{4}(1 +m\theta\bar\theta)\bar D\bar Z DZ -\frac{m^2}{4}\,\theta\bar\theta\,Z\bar Z
+ \frac12(1 - m\theta\bar\theta)\Pi\bar\Pi \Big]. \nonumber
\eea

\end{document}